\documentclass[12pt]{article}
  \usepackage{latexsym}
\else\oddsidemargin25mm\evensidemargin25mm\marginparwidth25mm\fi%
\begin{document}
%%%%%%%%%%%%%%%%%%%%%%%%%%%%%%%%%%%%%%%%%%
%%%%%%%%%%%%%%%%%%%%%%%%%%%%%%%%%%%%%%%%%%
% %%%%%%%%%%%%%%%%%%%%%%%%%%%%%%%%%%%%%%%%%
%%%%%% ANNAMACRO %%%%%%%%%%%%%%%%%%%%%%%%%%
%%%%%%%%%%%%%%%%%%%%%%%%%%%%%%%%%%%%%%%%%%%
%%%%%%%%%%%%%%%%%%%%%%%%%%%%%%%%%%%%%%%%%%%
\newcommand{\ft}[2]{{\textstyle\frac{#1}{#2}}}
\newcommand{\QED}{{\hspace*{\fill}\rule{2mm}{2mm}\linebreak}}
\def\dop{{\rm d}\hskip -1pt}
\def\bfone{\relax{\rm 1\kern-.35em 1}}
\def\bfzero{\relax{\rm I\kern-.18em 0}}
\def\inbar{\vrule height1.5ex width.4pt depth0pt}
\def\IC{\relax\,\hbox{$\inbar\kern-.3em{\rm C}$}}
\def\ID{\relax{\rm I\kern-.18em D}}
\def\IF{\relax{\rm I\kern-.18em F}}
\def\IK{\relax{\rm I\kern-.18em K}}
\def\IH{\relax{\rm I\kern-.18em H}}
\def\II{\relax{\rm I\kern-.17em I}}
\def\IN{\relax{\rm I\kern-.18em N}}
\def\IP{\relax{\rm I\kern-.18em P}}
\def\IQ{\relax\,\hbox{$\inbar\kern-.3em{\rm Q}$}}
\def\IR{\relax{\rm I\kern-.18em R}}
\def\IG{\relax\,\hbox{$\inbar\kern-.3em{\rm G}$}}
\font\cmss=cmss10 \font\cmsss=cmss10 at 7pt
\def\ZZ{\relax\ifmmode\mathchoice
{\hbox{\cmss Z\kern-.4em Z}}{\hbox{\cmss Z\kern-.4em Z}}
{\lower.9pt\hbox{\cmsss Z\kern-.4em Z}} {\lower1.2pt\hbox{\cmsss
Z\kern .4em Z}}\else{\cmss Z\kern-.4em Z}\fi}
\def\a{\alpha} \def\b{\beta} \def\d{\delta}
\def\e{\epsilon} \def\c{\gamma}
\def\G{\Gamma} \def\l{\lambda} \def\g{\gamma}
\def\L{\Lambda} \def\s{\sigma} \def\l{\lambda} \def\p{\psi} \def\pb{\overline{\psi}}
\def\cA{{\cal A}} \def\cB{{\cal B}}
\def\cC{{\cal C}} \def\cD{{\cal D}}
    \def\cF{{\cal F}} \def\cG{{\cal G}}
\def\cH{{\cal H}} \def\cI{{\cal I}}
\def\cJ{{\cal J}} \def\cK{{\cal K}}
\def\cL{{\cal L}} \def\cM{{\cal M}}
\def\cN{{\cal N}} \def\cO{{\cal O}}
\def\cP{{\cal P}} \def\cQ{{\cal Q}}
\def\cR{{\cal R}} \def\cV{{\cal V}}\def\cW{{\cal W}}
%
%
%%%%%%
%% This File is the titlepage %%%%%%%%%%%%%%%%%%%%%%%%%%%%%%%%
%%%%%%%%%%%%%%%%%%%%%%%%%%%%%%%%%%%%%%%%%%%%%%%%%%%%%%%%%%%%%
\begin{titlepage}
\hskip 5.5cm
\vbox{}
\hskip 2.5cm
\vbox{
\hbox{CERN-TH/2001-069}\hbox{March 2001}
}
\vfill
\begin{center} {\LARGE \bf On Fermion Masses,
Gradient Flows and Potential in Supersymmetric Theories}
\vskip 3cm {\bf Riccardo D'Auria$^\dagger$ and Sergio Ferrara$^\star$}\\
\skip 0.5cm
 {\it $^\dagger$ Dipartimento di Fisica, Politecnico
diTorino,Italy}\\ {\it Corso Duca degli Abruzzi 24, I-10129
Torino, Italy }\\ {\it INFN, Sezione di Torino, Italy.} \\{ \it
$^\star$ CERN, Theory Division, CH 1211 Geneva 23, Switzerland,
and } \\{\it Laboratori Nazionali di Frascati, INFN, Italy.}

\vskip 2cm

\begin{abstract}

In any low energy effective supergravity theory general
formulae exist which allow one to discuss fermion masses, the
scalar potential and breaking of symmetries in a model independent set up. A particular role in
this discussion is played by Killing vectors and Killing
prepotentials. We outline these relations in general and specify
then in the context of $N=1$ and $N=2$ supergravities in four
dimensions. Useful relations of gauged quaternionic geometry underlying
 hypermultiplets dynamics are discussed.
\end{abstract}

\end{center}
\end{titlepage}

\vfill\eject
\section{Introduction}

Supersymmetry Ward identities play a crucial role in disentangling general properties
of effective supersymmetric Lagrangians arising from some more fundamental theory
at the Planck scale where gravity is strongly coupled.

A particular role is played by supersymmetry relations on the scalar potential \cite{maiafe},\cite{cgp}
and on supersymmetry preserving vacua which are at the basis for the discussion.
of partial supersymmetry breaking \cite{b2},\cite{fgp},\cite{apt},\cite{m},\cite{tava} and of BPS configurations
and non perturbative string or M theory vacua,\cite{fks},\cite{ps},\cite{m},\cite{ma}.
 Some of these relations where studied
long ago,\cite{cgp}, but more recently a more careful analysis of supersymmetry preserving
configurations has played a crucial role in the study of the so
called "attractor mechanism" \cite{fk1}
for charged "black holes" in four and five dimensions \cite{fgk} as well as for the study of
supergravity flows \cite{kl},\cite{ckrrs} related to the so called
"renormalization group flow" \cite{gppz},\cite{pgpw}in the framework of the $AdS/CFT$ correspondence \cite{agmoo}.

Very recently these relations have been applied to a variety of interrelated problems
such as domain walls in five dimensional supergravity \cite{losw},\cite{bhlt},\cite{cd},\cite{gz},
\cite{bc},\cite{bgs}
 , and supergravity instantons \cite{gs}
responsible for non perturbative corrections to the hypermultiplet moduli space
in Calabi--Yau compactifications \cite{bbs}.

 In the present note we make some general
consideration on scalar potentials, fermion masses and Killing
prepotentials in a generic supersymmetric theory encompassing any
low-energy effective Lagrangian of a more fundamental theory
which at low energy incorporates a theory of gravity with
$N$-extended supersymmetry.
\noindent
Much of the information comes
from the analysis of simple terms in the supersymmetry variation
of the effective action, namely terms with one fermion and one
boson (or its first derivative).

It is shown that general formulae for fermion masses and scalar
potential exist which are simply related to the fermion shifts of
the supersymmetry transformation laws; in particular the $N=1$
and $N=2$ structures of the matter coupled supergravities
\cite{cfgv},\cite{bw1},\cite{b1},
\cite{dlv},\cite{a1}, can be
recovered in a simplified form. To illustrate the general
procedure we limit ourselves to the four-dimensional case, but it
is straightforward to see that our considerations can be extended
also to higher dimensions.
\noindent
In the particular case of $N=2$ supergravity with arbitrary gauge interactions turned on
an important role is played by gauged quaternionic \cite{bw},\cite{fs},\cite{dff} and special geometry
\cite{dlv}, \cite{s},\cite{cdf}
 which was discussed in full detail some time ago \cite{dff},\cite{a1}.
Here we are able to find new relations between Killing prepotentials which
allow us to show that some gradient flow relations
due to supersymmetry are merely due to some simple properties
of special and quaternionic geometry in presence of gauged isometries.
These relations purely depend on the geometrical data of the theory,
including gauging of isometries of the scalar manifold.
\footnote {A short account of the geometrical approach to Supergravity in the $N=2$ case
is given in \cite{a1}( see especially the Appendices of the second paper); a more
general reference is \cite{cdf1},Volume 2.}
This note is organized as follows: in section 2 we set up the formalism
and derive some basic relations between scalar potential, fermionic shifts
and fermion mass--matrices.

In section 3 and 4 we specify these relations to $N=1$ and $N=2$ theories
and recast some results in a model independent set up.

Section 4 is particularly relevant because it deals with $N=2$ supergravity
with general interactions of vector multiplets and hypermultiplets turned on \cite{a1}.
Here some interesting relations emerge due to the special structure
of coupled special and quaternionic geometries in presence of gauge isometries.

One of the amusing results, already noted in some special cases, is that the
(non derivative) part of the spin $\frac {1}{2}$--shifts can be written
in terms of the (covariant) derivative of some scalar functions \cite{bhlt},\cite{bgs},\cite{gs},\cite{kls}
 (the hypermultiplet and vector multiplet prepotentials)
exactly as in the case of charged (abelian) black--hole configurations \cite{fk1},
where the "central charge" matrix is here replaced by the $SU(2)$ valued
prepotential matrix.

An Appendix with the basic relations of gauged quaternionic geometry is included.

\section{The formalism: entangling supersymmetry with geometry}
\setcounter{equation}{0}
 We write down the generic form of $4D$
$N$ -extended supergravity theory (omitting four-fermion terms)in the following way:
\begin{eqnarray}
\label{lag} (\rm{det}V)^{-1}\mathcal{L}
&=& -\frac{1}{2}\mathcal{R}
+ {\rm i}\left( \bar {\cal N}_{\Lambda\Sigma}{\cal
F}^{-\Lambda}_{\mu\nu}{\cal F}^{-\Sigma \mu\nu} - {\cal
N}_{\Lambda\Sigma} {\cal F}^{+\Lambda}_{\mu\nu}{\cal F}^{+ \Sigma
{\mu\nu}}\right )  + \hat P^{IA}_{\mu} \hat P^{\mu}_{IA}
\nonumber\\
&+& {{\epsilon^{\mu\nu\lambda\sigma}}\over{\sqrt{-g}}}
\left( \bar\psi^A_\mu\gamma_\sigma D_\nu\psi_{A\lambda} -
\bar\psi_{A\mu} \gamma_\sigma D_\nu\psi^A_{\lambda} \right ) +
{{\rm i}\frac{1} {2}} \left(\bar\lambda^{I}\gamma^\mu
\nabla_\mu\lambda_I +\bar\lambda_{I} \gamma^\mu
\nabla_\mu\lambda^{I}\right )
\nonumber\\
&-& \hat P^{IA}_{\mu}
\bar \lambda_I \gamma^{\nu}\gamma^{\mu}\psi_{A \nu} - \hat
P_{IA\mu} \bar\lambda^I \gamma^{\nu}\gamma^{\mu}\psi_{\nu}^A +
 \nonumber \\
&+&{\cal F}^{\Lambda}_{\mu\nu}{\cal N}_{\Lambda\Sigma}\left
(L^\Sigma_{AB}\overline{\psi}^{\mu A}\psi^{\nu B}+L^\Sigma_{IA}\overline{\psi}^{\mu A}\gamma^\nu
\lambda^{I}+ L^{\Sigma}_{IJ}\overline{\lambda}^I \gamma^{\mu \nu}\lambda^J + h.c.\right)\nonumber \\
&+& 2 S_{AB}\bar \psi_{\mu}^A \gamma^{\mu\nu} \psi_{\nu}^B + 2 {\overline S}^{AB}\bar\psi_{\mu A} \gamma^{\mu\nu}
 \psi_{\nu  B} \nonumber \\
&+& {\rm{i}} N_I^A {\bar \lambda}^I \gamma^{\mu} \psi_{\mu A}
+{\rm{i}} N^I_A {\bar \lambda}_I \gamma^{\mu} \psi_{\mu}^A +
{\mathcal M}^{IJ} {\bar \lambda}_I \lambda_J
+{\mathcal M}_{IJ} \bar\lambda^I\lambda^J - \mathcal{V}(q)
\end{eqnarray}
\noindent
 where $q^u$,$\lambda_I$,$\psi_{A \mu}$ are the scalar
fields, the spin $\frac{1} {2}$ fermions, and $\psi _{A \mu}$ the
gravitino fields.The labels of the fields are as follows;
 we indicate by $A,B,\dots $ the indices of the fundamental representation of the
$R$-symmetry group $SU(N)$ $\otimes U(1)$, their lower (upper)
position indicating their left (right) chirality.
Actually the indices $I$ besides to enumerate the
fields, are a condensed notation which encompasses various
possibilities; if the fermions belong to vector multiplets we have
to set $I\rightarrow IA$ since they also transform under
R-symmetry ; if they refer to fermions of the gravitational
multiplet they are a set of three $SU(N)$ antisymmetric indices:
$I \rightarrow [ABC] $. (In the particular case of $n_H$ hypermultiplets
$I\rightarrow \alpha$ where $\alpha$ is in the fundamental of
$Sp(2n_H)$ ).

The matrices entering the Lagrangian are all dependent on the
scalar fieds $q^i$. $\mathcal {N}_{\Lambda\Sigma}$ is the
symmetric matrix of the vector field-strengths, with
$\Lambda,\Sigma$ indices in the symplectic representation under
which they transform; $S_{AB},N^I_A,M^{IJ}$, together with their
hermitian conjugates ${\bar S}^{AB},N_I^A,M_{IJ}$,are matrices of order
$g$ in the gauge coupling constant while the scalar potential
$\mathcal {V}(q)$ is of order $g^2$. Note that ${\bar S}^{AB},M^{IJ}$ are
the mass matrices of the gravitino and the spin $\frac{1} {2}$
fermions.The structures  $ L^\Sigma_{AB},\,L^\Sigma_{IA},\,
 L^{\Sigma}_{IJ}$ are coset representatives of the $\sigma$-model $G/H$ for $N>2$
 while they are special Geometry objects for $N=2$ and,  for $N=1$they are related to the kinetic
 matrix of the vectors (with $L^\Sigma_{AB}=0$).
  Finally we note that the covariant derivatives acting on the spinors contain
besides the spin connection the composite connections of the $\sigma $-model of the scalar field
$q^u$. Furthermore in presence of gauging the composite connections are gauged according to the usual procedure
for $N>2$ theories where the $\sigma $- model is a coset ; their explicit expressions
for the case of $N=1$ and $N=2$ gauged supergravity are given in the following. In particular
$\hat P^{I A}$ are the the gauged vielbein
1-forms of the scalar manifold defined as
\begin{equation}
\hat P_{IA\mu} =  P_{IAu}\left(\partial_{\mu} q^u + g
A^{\Lambda}_{\mu} k^u_{\Lambda}\right)
\end{equation}
\noindent
 where $P_{I Au}$ is the ordinary vielbein of the scalar
manifold. Also in this case the index $I$ of the vielbein must be
given the same interpretation as explained in the case of the spin
$\frac{1}{2}$ fields.~\ Moreover for any boson field $v$ carrying
$SU(N)$ indices we have that lower and upper indices are related
by complex conjugation,namely:
\begin{equation}\label{cc}
(v_{AB\dots})^* \sim {\bar v}^{AB \dots}\equiv v^{AB}
\end{equation}
 When $N>2$, so that
the scalar manifold is a coset $G/H$, the gauged vielbein 1-form
can be rewritten
\begin{equation} \label{vielb}
\hat P_{I A}= {\left (L^{-1}\,dL \right)}_{I A} + g
{L^{-1}}_{I\Gamma} A^{\Lambda} (T_{\Lambda})^{\Gamma}_{\, \Pi}
L^{\Pi}_{ A}
\end{equation}
due to the general relation
\begin{equation}\label{rel}
P_{I Au} k^u_{\Lambda} \equiv \left(L^{-1}\partial_u L
\right)_{I A} k^u_{\Lambda}  =L^{-1}_{I\Gamma}
(T_{\Lambda})^{\Gamma}_{\,\,\Pi} L^{\Pi}_{A}
\end{equation}
\noindent
 where $T_{\Lambda}$ are the generators of the gauge
group.

We now write down the relevant terms of the supersymmetry
transformation laws of the various fields in order to perform the
supersymmetry variation of the Lagrangian; this will allow us to identify
the differential equations for the fermionic shifts and other important relations
between geometrical quantities mentioned in the introduction. We have:
\begin{eqnarray}
\label{trapsi}\delta \psi_{A \mu} &=& \nabla_{\mu}\varepsilon_A
+ T_{AB\mu\nu}^{(-)}\gamma^\nu \varepsilon^B
+S_{AB}\gamma _\mu \varepsilon ^B + \cdots\\
\label{tralamb}\delta \lambda_I &=& {\rm{i}} \hat P_{uIA}\gamma_\mu
\varepsilon^A
\nabla_{\mu} q^u + T_{I\mu\nu}^{(-)} +N_{I}^{A}\varepsilon_A+ \cdots \\
\label{traviel}\delta V^a_{\mu} &=& -\rm{i} {\overline \psi}_A
\gamma_{\mu} \varepsilon^A  -\rm{i} {\overline \psi}^A
\gamma_{\mu} \varepsilon_A -
\\
\label{travec}\delta A^{\Lambda}_{\mu} &=& 2 f^{\Lambda [A B]}
{\overline \psi}_{A \mu} \varepsilon_B +{\rm{i}} f^{\Lambda A}_I
\bar \lambda^I \gamma_{\mu}
\varepsilon_A \\ \nonumber&+& 2 f^{\Lambda}_{[A B]}
{\overline \psi}^{A}_{\mu} \varepsilon^B +{\rm{i}} f^{\Lambda I}_{A}
\bar \lambda_I \gamma_{\mu}
\varepsilon^A \\
 \label{trasca}\delta q^u P_{uIA}&=& \bar \lambda_I
\varepsilon_A \\
\label{trascabis}\delta q^u P_{u}^{IA}&=& \bar \lambda^I
\varepsilon^A
\end{eqnarray}
where we have written the transformations only for the left-handed fermions
and the dots denote trilinear fermion terms In the previous formulae
$ f^{\Lambda [A B]}$ and $f^{\Lambda A}_I$ are related to the symplectic
sections on the scalar manifold of the vector multiplets for $N>1$,
while for $N=1$ the former is identically zero and the latter reduces to a Kronecker delta.
Moreover
$T^{(-)}_{AB\mu \nu} ={\rm i}\bar {f}^{-1}_{AB\Lambda} F^{(-)\Lambda}_{\mu \nu}$
and $T^{(-)}_{I\mu \nu} ={\rm i}\bar {f}^{-1}_{I\Lambda} F^{(-)\Lambda}_{\mu \nu}$
are the "dressed" field strengths of the graviphotons and of the vector multiplets; of course
since in $N=1$ $f^{\Lambda [A B]}\equiv 0$ and $f^{\Lambda A}_I = \delta ^\Lambda_I$
($A=1$ and $\Lambda$ index of the vector multiplet),  these terms are absent in the
transformatiom laws
of the gravitino and of the spinor of the chiral multiplet in the $N=1$ theory .\\
 We are going to explore the
invariance of (\ref{lag}) (up to a total derivative) for terms of
the form $f\,\varepsilon\,B $ where $f$ is a fermion and $B$ is a
function of the scalar fields.

We have to look to two kinds of terms:
\begin{itemize}
\item terms with one derivative
\item terms with no derivatives
\end{itemize}

In the first case we can choose $f\, \partial\varepsilon\,q $ and
$f\, \varepsilon\,\partial q $ as independent variations \cite{maiafe} since all
these terms are independent. It is a simple exercise , first
carried out in \cite{maiafe}, to see that the terms containing the
derivative of the supersymmetry parameter just fix the couplings
$2\,{\bar S^{AB}}{\overline \psi}_{\mu}^A\gamma^{\mu\nu}\psi_{\nu}^B
+{\rm{i}}N_I^A \bar\lambda^I \gamma^{\mu}\psi_{\mu A} +h.c.$ of
the Lagrangian in terms of the shifts proportional to $g$ of the
equations (\ref{trapsi}),(\ref{tralamb}).

The terms with no derivatives of the form $f\,\varepsilon\,B $
instead give rise to two important relations \cite{maiafe}. The first one is
due specifically to the terms $ \psi_A \gamma_{\mu} \varepsilon^A
\,B(q) $ which determine the scalar potential $V(q)$ in terms of
the squared modulus of the shifts (\ref{trapsi}),(\ref{tralamb});
one finds:
\begin{equation}\label{pot}
 \delta_B^A \,V(q) = -12 \bar S^{AC}\,S_{CB} + N_I^A\,N^I_B
\end{equation}

The second one is due to terms of the form $\lambda \varepsilon
B(q)$ and  their h.c.~\ that give rise to a formula from which the
Goldstone theorem for supergravity can be derived:
\begin{equation} \label{goldstone}
\frac {\partial V}{\partial q^u} P_{IAu} =4{\rm{i}} N^{B}_I\, S_{BA} + 2
{\mathcal {M}}_{IJ}\,N^{J}_{A}
\end{equation}

Tracing equations (\ref{pot}) with respect to $A,B$ and
differentiating with respect to $q^u$, by comparison with equation
(\ref{goldstone}), it follows that there must be some relation
between $N_I^A$ and $\partial S_{AB}$ as well as between
${\mathcal {M}}^{IJ}$ and $\partial N^J_A$. We shall refer to
these relations as "gradient flows" for the fermionic shifts.
These gradient flows can be obtained in the simplest way by
looking at the terms $f\, \varepsilon\,\partial q $ of the first
item ( terms with one derivative) which have not been yet
considered.

Let us first consider the the equations derived when considering
terms of the form $\psi\,\varepsilon\,\partial q$. There are two
independent structures proportional to the currents with
$\delta^{\mu \nu}$ and $\gamma^{\mu \nu}$,respectively. The
$\delta^{\mu \nu}$ current gives the equation:
\begin{equation}\label{fk}
 k^u_{\Lambda}\,f^{\Lambda\,[A B]} + N_I^{[A}\,P^{B]I\,u} = 0
\end{equation}
where one has to take into account the contribution coming from
the kinetic term of the scalars due to the definition of $\hat
P^{I A}_u$ which contributes through $\delta A^{\Lambda}_{\mu}$ given in
equation (\ref{travec}). Equation (\ref{fk}) relates the Killing
vector $ k^u_{\Lambda} $ ~\ $(\delta q^u = \xi ^{\Lambda}
\,k^u_{\Lambda})$\, to the spin $\frac {1}{2}$ shifts $N_I^{A}$.

 The terms proportional to the $\gamma^{\mu \nu}$ current
yield the gradient flow:
\begin{equation}\label{gfS}
D_u S^{AB} = N^{(A}_I\,P^{B) I}_u
\end{equation}

Considering next the equation coming from the terms $\lambda
\varepsilon \partial q$ we find the gradient flow of the spin
$\frac {1}{2}$ shifts :
\begin{equation}\label{gfN}
 \nabla_u N_I^A = g_{uv} \,k^v_{\Lambda}\,f^{\Lambda A}_I + 2\,
 P_{I B\,u}\,S^{BA}\,+2\, {\mathcal {M}}_{IJ} \,P^{J A}_u
 \end{equation}

Alternatively eq.s (\ref{fk}), (\ref{gfS}) can be cast in the
following form:
\begin{equation}\label{cast}
D_u S^{AB} =P^{A I}_u N^{B}_I - k_{u\,\Lambda}\,f^{\Lambda\,[A
B]}
\end{equation}
which is analogous to eq. (\ref{gfN}).

We note that the previous results determine the full fermionic
mass matrix ${\mathcal{M}}_{IJ}$ through eq. (\ref{gfN}).

If there are multiplets with no scalars as it happens in the
$N=1$ case then the fermionic mass matrix for the fermions of such
multiplets is obtained by looking in the variation of the
Lagrangian to extra terms of the form $\lambda \varepsilon
\,\mathcal{F}$, $\mathcal{F}$ being the field-strength of the
vector replacing in this case the $\partial q$ factor: indeed if
$\lambda$ has no scalar partner it must certainly have a vector
partner and the mass matrix of the fermions can be obtained by
the aforementioned variation.

In a different fashion also behave multiplets where the fermions
are the only partner of scalar fields (Wess-Zumino multiplets in
$N=1$ and hypermultiplets in $N=2$) because in those cases
$f^{\Lambda [A B]}$ and $ f^{\Lambda I A}$ do not exist in the eq.
(\ref{travec}) and therefore they do not enter in the
determination of $\nabla_u N^A_I$. Under these circumstances
$\nabla_u N^A_I$ and ${\mathcal{M}}_{IJ}$ can be expressed through
eq.s (\ref{gfN}), (\ref{cast}) purely in terms of the gravitino
mass matrix $S_{AB}$ and its derivatives \cite{cfgv}.

\section{The N=1 case}
\setcounter{equation}{0}
In order to apply our formulae to the $N=1$ case we recall that
the scalar manifold is in this case a  K\"{a}hler-Hodge manifold \cite{bw}
and that the R- symmetry reduces simply to $U(1)$.~\ It is
convenient in this case to use as "vielbeins" the differential of
the complex coordinates $dz^i,d \bar z^{i^\star}$ where $z^i(x)$
are the complex scalar fields parametrizing the K\"{a}hler-Hodge
manifold of (complex) dimension $n_C$. Therefore in the present
case we have to set $q^u \rightarrow (z^i,\bar z^{i^\star})$. The
spin $\frac {1}{2}$ fermions are either in chiral or in vector
multiplets. So the index $I$ runs over the number $n_V +n_C$ of
vector and chiral multiplets, $I =1, \dots,n_V + n_C$. It is
convenient to assign the index $\Lambda$, the same as for the
vectors, to the fermions of the vector multiplets and we will
denote them as $\lambda^{\Lambda}$,\, $\Lambda =1,\dots,n_V$;
the fermions of the chiral multiplets will instead be denoted by
$\chi^i,\chi^{i^\star}$ in the case of left-handed or
right-handed spinors, respectively. Since the gravitino and the
gaugino fermions have no $SU(N)$ indices their chirality will be
denoted by a lower or an upper dot for left-handed or right
handed fermions respectively, namely {$(\psi_{\bullet}$,
$\psi^{\bullet}$)}; {($ \lambda_{\bullet}^{ \Lambda}$,
$\lambda^{\bullet\Lambda})$}. Note that we have two metrics,
namely the K\"{a}hler metric $g_{ij^\star}$ of the scalar
manifold and the metric ${\mathcal {N}}_{\Lambda \Sigma}$ of the vector kinetic term with
symplectic indices $ \Lambda, \Sigma $. In writing covariant derivatives we use the symbol
$\nabla_\mu$ which takes into account covariance under Lorentz, gauge, $\sigma$-model
 reparametrizations and furthermore $U(1)$ K\"{a}hler transformations. Indeed  since the scalar manifold is
  a K\"{a}hler-Hodge
manifold all the fields and the bosonic sections have a definite
$U(1)$ weight $p$ under $U(1)$. We have
\begin{eqnarray}\label{weight}
  p(V^a_{\mu})&=& p(A^{\Lambda})= p(z^i)= p(g_{ij^\star})=p(
  {\mathcal {N}}_{\Lambda \Sigma}) = p( D^{\Lambda}) = p(P_{\Lambda}) =p({\mathcal {V}}) =0 \nonumber \\
  p(\psi_{\bullet})&=& p(\chi^{i^\star})=
  p(\lambda^{\Lambda}_{\bullet})=p(\varepsilon_{\bullet}) = \frac {1} {2} \nonumber \\
p(\psi^{\bullet})&=& p(\chi^{ i})=
  p(\lambda^{\Lambda \bullet})=p(\varepsilon^{\bullet}) = - \frac {1} {2} \nonumber \\
p(L)&=& p( {\mathcal {M}}_{ij}) = p( \overline {\mathcal {M}}_{\Lambda
\Sigma}) =1 \nonumber \\
p(\bar L)&=& p({\overline {\mathcal M}}_{i^\star j^\star}) =
p( {\mathcal {M}}_{\Lambda \Sigma}) =- 1
\end{eqnarray}

Accordingly, when a covariant derivative acts on a field  $\Phi$ of
weight $p$ it is  also $U(1)$ covariant (besides possibly Lorentz,
gauge and scalar manifold coordinate symmetries) according to the
following definitions:
\begin{equation}
\begin{array}{ccccccc}
\nabla_i \Phi &=&
 (\partial_i + {1\over 2} p \partial_i {\cal K}) \Phi &; &
\nabla_{i^*}\Phi &=&(\partial_{i^*}-{1\over 2} p \partial_{i^*}
{\cal K}) \Phi \cr
\end{array}
\end{equation}
where $ {\mathcal {K}(z,\bar z)}$ is the K\"{a}hler potential.
\noindent
A covariantly holomorphic section of is defined by the
equation: $ \nabla_{i^*} \Phi = 0  $.\\
Denoting the Lorentz covariant derivative on a generic spinor
\begin{equation}\label{lorentz}
\mathcal D_\mu \phi = \partial_\mu \phi - \frac {1} {4}\omega^{ab} \gamma_{ab}\phi
\end{equation}
the complete covariant derivatives on the chiral supersymmetry parameter $\varepsilon_\bullet$,
 the spinors $\lambda^\Lambda\bullet$ and
$\chi^i$ are then defined as follows:
\begin{eqnarray}
\nabla_\mu \varepsilon_\bullet&=&\mathcal D_\mu \varepsilon_\bullet +{\rm i}p \hat Q_\mu \varepsilon_\bullet \\
\nabla_\mu \lambda^\Lambda_\bullet& =&\mathcal D_\mu \lambda^\Lambda_\bullet +{\rm i}p \hat Q_\mu \lambda^\Lambda_\bullet \\
\nabla_\mu \chi^i&=&\mathcal D_\mu \chi +{\rm i}p \hat Q_\mu \chi^i +\hat {\Gamma}^i_{jk}\partial_\mu z^k\chi^j
\end{eqnarray}
where the 1-forms $\hat {Q}$ and ${\hat \Gamma}^i_j$ are defined as follows:
\begin{eqnarray}
\hat Q &=& Q +A^\Lambda P_\Lambda \nonumber \\
Q &=&- \frac {i} {2}\left (\partial_i K dz^i-\partial_{i^*} K dz^{i^*}\right )\\
\hat \Gamma^i_j& =& \Gamma^i_j + \partial_j k^i_\Lambda A^\Lambda
\end{eqnarray}
Moreover we have:
\begin{equation}\label{dev}
  \nabla_\mu z^i= \partial_\mu z^i + k^i_\Lambda A^\Lambda_\mu
\end{equation}

In these equations $k^i_\Lambda(z^i)$ and $P_\Lambda(z^i,z^{i^*})$ are the Killing vectors
of the gauged isometries of the K\"{a}hler-Hodge manifolds and its associated prepotential.
In K\"ahler geometry $k^i_{\Lambda}
\,=\,k^i_{\Lambda}(z)$ is
 holomorphic and satisfies the following relations:

\begin{eqnarray}
\nabla_i k_{j \Lambda}&=&0\\
\nabla_i k_{j^\star \Lambda}&=& \nabla_{j^\star} k_{i \Lambda}
\end{eqnarray}
Its relation with the prepotential is:
\begin{equation}\label{kill}
  k^i_{\Lambda} = {\rm {i}}g^{ij^\star} \partial_{j^\star}
P_{\Lambda}
\end{equation}

 Using these conventions we have the following
 $N=1$ supergravity
Lagrangian (up to 4-fermion terms) :
\begin{eqnarray}
\label{lag1}(\rm{det}V)^{-1}\mathcal{L}&=& -\frac{1}{2}\mathcal{R}
+ {\rm i}\left( {\overline {\cal N}}_{\Lambda\Sigma}{\cal
F}^{-\Lambda}_{\mu\nu}{\cal F}^{-\Sigma \mu\nu} - {\cal
N}_{\Lambda\Sigma} {\cal F}^{+\Lambda}_{\mu\nu}{\cal F}^{+ \Sigma
{\mu\nu}}\right )  + g_{ij^\star}\nabla_{\mu}z^i
\nabla^{\mu}z^{j^\star} \nonumber \\&+&
{{\epsilon^{\mu\nu\lambda\sigma}}\over{\sqrt{-g}}} \left(
\bar\psi^{\bullet}_\mu\gamma_\sigma D_\nu\psi_{\bullet \lambda} -
\bar\psi_{\bullet \mu} \gamma_\sigma D_\nu\psi^{\bullet}_{\lambda}
\right ) +\frac {1}{8} \left({\cal
N}_{\Lambda\Sigma}\bar\lambda^{\bullet \Lambda}\gamma^\mu
\nabla_\mu\lambda_{\bullet}^{\Sigma} -{\overline {\cal
N}}_{\Lambda\Sigma}\bar\lambda_{\bullet}^{ \Lambda} \gamma^\mu
\nabla_\mu\lambda^{\bullet \Sigma} \right) \nonumber \\&-&
{\rm{i}} \frac{1}{2} g_{ij^\star} \left(\bar\chi^{i}\gamma^{\mu}
\nabla_{\mu} \chi^{j^\star} +\bar \chi^{j^\star} \gamma^{\mu}
\nabla_{\mu} \chi^{i} \right )- g_{ij^\star} \left(
 \bar \psi_{\bullet \nu} \gamma^{\mu}\gamma^{\nu} \chi^i \nabla^{\mu}\bar z^{\bar j}
+ \bar \psi^{\bullet}_{ \nu} \gamma^{\mu} \gamma^{\nu} \chi^{\bar
 j} \nabla_{\mu}z^{i} \right) \nonumber \\
 &+&  \,{\rm{i}}\, {\rm Im {\cal N}_{\Lambda\Sigma}} \left ( {\cal
 F}^{+\Lambda}_{\mu\nu}\bar \lambda_{\bullet}^{ \Sigma}
 \gamma^{\mu} \psi ^{\bullet \nu} + {\cal
 F}^{-\Lambda}_{\mu\nu}\bar \lambda^{\bullet \Sigma}
 \gamma^{\mu} \psi _{\bullet}^{ \nu} \right)\nonumber \\
&-&\frac {\rm {i}}{8} \left (\partial_i {\overline {\mathcal
{N}}}_{\Lambda \Sigma} {\cal F}^{-
\Lambda}_{\mu\nu} \bar \chi^i
 \gamma^{\mu \nu} \lambda _{\bullet}^{\Sigma} - (\partial_{i^\star} {\mathcal  {N}}_{\Lambda \Sigma} {\cal
 F}^{+\Lambda}_{\mu\nu} \bar \chi^{i^\star}
 \gamma^{\mu \nu} \lambda ^{\bullet \Sigma} \right) \nonumber
 \\&+&2L\bar\psi_{\mu}^\bullet\gamma^{\mu\nu}\psi_{\nu}^\bullet+2\bar L
\bar\psi_{\mu \bullet}\gamma^{\mu\nu}\psi_{\nu \bullet}\nonumber \\
&+&{\rm{i}}
 g_{ij^\star} \left ( {\overline N}^{\bar j} \bar\chi^i
\gamma^{\mu}\psi_{\mu}^{\bullet}+N^i \bar\chi^{j^\star}
\gamma^{\mu}\psi_{\bullet \mu}\right) + \frac{1}{2} P_{\Lambda}
\left( \bar \lambda^{\bullet \Lambda} \gamma^{\mu} \psi _{\bullet
\mu}- \bar \lambda_{\bullet} ^{\Lambda} \gamma^{\mu} \psi
_{\mu}^{\bullet} \right)\nonumber \\ &+&\mathcal M_{ij}
\bar\chi^i\chi^j + {\overline \mathcal M}_{i^\star j^\star}
\bar\chi^{i^\star} \chi^{j^\star} +\mathcal{M}_{\Lambda \Sigma}
\bar\lambda^{\Lambda}_{\bullet}\lambda^{\Sigma}_{\bullet}+{\overline
{\mathcal{M}}}_{\Lambda \Sigma} \bar\lambda^{\Lambda
\bullet}\lambda^{\Sigma \bullet} \nonumber \\ &+& +
+\mathcal{M}_{\Lambda i} \bar\lambda^{\Lambda}_{\bullet} \chi ^i
+{\overline{\mathcal{M}}}_{\Lambda i^\star} \bar\lambda^{\Lambda
\bullet} \chi ^{i^\star} \nonumber \\ &-& {\mathcal {V}}(z,\bar z, q)
\end{eqnarray}
where we have defined
\begin{eqnarray}\label{self}
{\mathcal {F}}_{\mu\nu}^{(\mp)\Lambda}\,&=&\,\frac{1}{2}\left({\mathcal {F}}_{\mu\nu}^{\Lambda}\,\mp\,
\rm {i} \star{\mathcal {F}^{\Lambda}}_{\mu\nu}\right)\nonumber \\
\star{\mathcal {F}^{\Lambda}}_{\mu\nu} &\equiv& \frac{1}{2} \epsilon_{\mu\nu\rho\sigma}
{\mathcal {F}}^{\rho\sigma\Lambda}\nonumber \\
\star{\mathcal {F}^{\Lambda (\pm)}}_{\mu\nu}\,&=&\,\mp \rm {i}{\mathcal {F}^{\Lambda (\pm)}}_{\mu\nu}
\end{eqnarray}
The Lagrangian is invariant under the following supersymmetry transformation laws for the
fields \cite{cfgv}, \cite{b1}:
\begin{eqnarray}
\label{trapsi1}\delta \psi_{\bullet \mu} &=& {\nabla}_{\mu}
\varepsilon_{\bullet}
+{\rm {i}} L(z, \bar z) \gamma_{\mu} \varepsilon^{\bullet} + \cdots\\
\label{trachi1} \delta \chi^i &=& {\rm {i}} \nabla_{\mu}z^i
\gamma^{\mu} \varepsilon_{\bullet}  + N^{i}\varepsilon_{\bullet}+ \cdots\\
\label{tralamb1}\delta \lambda^{\Lambda}_{\bullet} &=& {\mathcal {F}}_{\mu
\nu}^{(-) \Lambda} \gamma^{\mu \nu} \varepsilon_{\bullet }
+{\rm {i}} D^{\Lambda} \varepsilon_{\bullet}+ \dots \\
\label{traviel1}\delta V^a_{\mu} &=& -\rm {i}
\psi_{\bullet} \gamma_{\mu} \varepsilon^{\bullet} + h.c.\\
\label{travec1}\delta A^{\Lambda}_{\mu} &=& \rm{i}\frac {1}{2}
\bar \lambda^{\Lambda}_{\bullet} \gamma_{\mu}
\varepsilon^{\bullet} + h.c.\\
 \label{trasca1}\delta z^i  &=& \bar \chi^i
\varepsilon_{\bullet}
\end{eqnarray}
\noindent
and the kinetic matrix ${\overline {\mathcal
{N}}}_{\Lambda \Sigma}$ turns out by supersymmetry to be a holomorphic function of $z^i$:
${\overline {\mathcal{N}}}_{\Lambda \Sigma}\,=\,{\overline {\mathcal
{N}}}_{\Lambda \Sigma}(z^i)\rightarrow {\mathcal
{N}}_{\Lambda \Sigma}\,=\,{\mathcal
{N}}_{\Lambda \Sigma}({\bar z}^{i^\star})$.\\
 Supersymmetry implies that all the quantities entering the
 transformation laws and the Lagrangian can be expressed in terms
 of the following geometric quantities: the covariantly holomorphic gravitino mass-matrix $L(z, \bar
 z)$ , the Killing vector real prepotential $P_{\Lambda}(z, \bar
 z)$ the K\"{a}hler potential and the holomorphic matrix $
 \overline{\mathcal {N}}_{\Lambda \Sigma}(z)$.\\
Indeed we have the following
 relations \footnote{For constant scalar background unbroken supersymmetry
 requires $\nabla_i\,L\,=\,0$ , i.e. the "norm"  ${\|L\|}^2\,=\,L\,\bar L$ to be extremized
 on the K\"ahler-Hodge manifold. This is the $N=1$ example of "attractor equation"
 \cite{fks},\cite{fk1}.}:
\begin{eqnarray}\label{def1}
L(z,\bar z)&=& W(z)e^{\frac {1}{2} {\mathcal {K}(z, \bar z)}} \\
\nabla_i {\overline L}(z^i,z^{i^*})&=& 0 \\
N^i &=& 2 g^{ij^\star} \nabla_{j^\star} \bar L \\
 \label{dlambda} D^{\Lambda} &=& 2 {\rm {Im}} ({\mathcal{N}}_{\Lambda
\Sigma})^{-1}
P_{\Sigma} \\
\label{def4} {\mathcal {M}}_{ij} &=& \nabla_i \nabla_j L \\
{\mathcal {M}}_{\Lambda \Sigma} &=& \frac {1} {4} N^i \partial_i
Im{{\mathcal{N}}}_{\Lambda
\Sigma} \\
 M_{\Lambda i}&=& - {\rm {i}} \frac {1}{4} {\rm {Im}}
{\mathcal{N}}_{\Lambda \Sigma} \partial_i D^{\Sigma} -\frac {1}
{2} k^{j^\star}_{\Lambda} g_{ij^\star} \label{mista}
 \\
{\mathcal {V}}&=& 4 \left ( -3 L\bar L  +\,g^{ij^\star}
\nabla_i\,L \nabla_{j^\star}  \bar L \,- \frac {1}{16} {\rm {Im}}
{\mathcal {N}}_{\Lambda \Sigma} D^{\Lambda } D^{\Sigma} \right)
\end{eqnarray}

Note that eq.s (\ref{dlambda}) and (\ref{mista}) imply:
\begin{equation}\label{eleg}
M_{\Lambda i}D^\Lambda\,=\,- {\rm {i}} \frac {1}{4} \partial_i({\rm {Im}}
{\mathcal{N}}_{\Lambda \Sigma} D^\Lambda D^{\Sigma}).
\end{equation}

Finally the gradient flows are :
\begin{eqnarray}
\nabla_i N^j &=& 2\,\delta^j_i \\
 \nabla_i \bar N^{j^\star}g_{kj^\star} &=& 2 {\mathcal {M}}_{ik} \\
\nabla_i  L &=& \frac {1}{2} g_{ij^\star} \bar N^{j^\star} \\
\nabla_{i^\star} L &=& 0
\end{eqnarray}
\section{The N=2 case}
\setcounter{equation}{0}
For the $N=2$ supergravity the scalar manifold is a product
manifold \cite{dlv}, \cite{dff}, \cite{cgf}
\begin{equation}\label{mani}
  {\mathcal M}^{(scalar)} = {\mathcal M}^{(vec)}\otimes {\mathcal M}^{(hyper)}
\end{equation}
since we have two kinds of matter multiplets, the vector
multiplets and the hypermultiplets. The geometry of ${\mathcal
M}^{(vec)}$ is described by the ${\it Special
\,\,K\ddot{a}hler\,\, geometry} $ \cite{dlv}, \cite{s}, \cite{cdf},
\cite{crtv} while the geometry of ${\mathcal
M}^{(hyper)}$ is described by $\it {Quaternionic\,\, geometry}$ \cite {al}, \cite{ga},
\cite{bw},\cite{dlv},\cite{dff},\cite{goi},\cite{dkv},\cite{dv}. A
full account of  ${\it Special
\,\,K\ddot{a}hler\,\, geometry} $ is given in ref.s \cite {a1}. As far as
$\it {Quaternionic\,\, geometry}$ is concerned, we have set an Appendix to the present paper since
the account given in the reference \cite {a1} and also in \cite{bw}, \cite{dff} do not comprise
several new important identities that we present in the Appendix.\\
With
respect to the general case we now have
\begin{equation}\label{conv}
 \Lambda = 1,\dots,n_V;\,\,A,B = 1,2;\,\,
 i=1,\dots,4n_H+2n_V;\,\,I=1,\dots n_H+n_V
\end{equation}
We will use the same notations and conventions as in ref
\cite{a1} where the complete theory of the $N=2$ supergravity has been
fully worked out in a geometrical setting.
\par
 Let us  now shortly describe how our
general framework particularizes to the present case.

 As in the case of $N=1$ we denote the complex scalars
parametrizing ${\mathcal M}^{(vec)}$ by {${z^i, \bar z^{\bar
i}}$}, while the scalars parametrizing ${\mathcal M}^{(hyper)}$
will be denoted by {$q^u$}. As already noted in the previous
section, when the index $I$ runs over the vector multiplets it
must be substituted by $IB$ in all the formulae relevant to the
vector multiplet, since the fermions $\lambda^{IA}$ are in the
fundamental of the $R$-symmetry group $U(2)$. Furthermore if we
use coordinate indices as in the $N=1$ case so that the vielbeins
of ${\mathcal M}^{(vec)}$ are simply $dz^i,d \bar z^{\bar i}$ we
have to perform the following substitutions:
\begin{eqnarray} \label{vielspe}
&&P^{IA}_u dq^u \rightarrow P^{I \,BA}_i dz^i =
-\epsilon^{AB} dz^i \nonumber \\
&&P^{I^\star A}_i dq^u \rightarrow  P^{I^\star\,BA}_{i^\star}
d\bar z^{i^\star} = -\epsilon^{AB} d \bar z^{i^\star}
\end{eqnarray}

In particular, the general objects $f^{\Lambda [A B]},\,
f^{\Lambda I A}$ introduced in equation \ref {travec} become in
our case:
\begin{equation}\label{cosrep}
f^{\Lambda [A B]}\,=\, \epsilon^{AB}\overline
L^{\Lambda};\,\,\,\,f^{\Lambda  A}_I \, \rightarrow f^{\Lambda
A}_{IB} =\delta ^A_B\,\nabla_{i^\star} \bar L^{\Lambda}
\end{equation}
 where $L^{\Lambda}(z, \, \bar z)$ and its "magnetic" counterpart $M_{\Lambda}(z, \, \bar
 z)= {\mathcal N}_{\Lambda\,\Sigma}\,L^{\Sigma}$
actually form a $2n_V$ dimensional covariantly holomorphic
section $V= (L^{\Lambda}, \,M_{\Lambda})$ of a flat symplectic
bundle.

When the index $I$ runs over the hypermultiplets we will rename
them as follows: $(I,J) \rightarrow (\alpha,\, \beta) $ and since
there are no vectors in the hypermultiplets we have $f^{\Lambda A}
_{\alpha} =0$

The vielbeins of the quaternionic manifold ${\mathcal
M}^{(hyper)}$ will be denoted by ${\mathcal U}^{\alpha\,A} \equiv
{\mathcal U}^{\alpha\,A}_u dq^u$ where $\alpha = 1,\dots ,2n_H$
is an index labelling the fundamental representation of
$Sp(2n_H)$. The inverse matrix vielbein is ${\mathcal
U}^{u}_{\alpha\,A}$. We raise and lower the indices $\alpha
,\beta, \dots$ and $A,B, \dots$ with the symplectic matrices
$C_{\alpha\,\beta}$ and $ \epsilon_{A\,B}$ according to the
following conventions

\begin{equation}
\epsilon^{AB}\,\epsilon _{BC} \,=\,-\, \delta^A _C ; \qquad
\qquad \epsilon^{AB}\,=\,-\,   \epsilon^{BA} \label{epsi}
\end{equation}
 \begin{equation}
\IC^{\alpha\beta}\,\IC _{\beta\gamma} \,=\,-\, \delta^\alpha _
\gamma ; \qquad  \qquad \IC^{\alpha\beta}\,=\,-\,
\IC^{\beta\alpha};  \label{cpsi}
\end{equation}
For any $SU(2)$ vector $P_A$ we have:
\begin{equation}
\epsilon _{AB}\, P^B \,=\,P_A ; \qquad  \qquad  \epsilon^{AB}\, P_B \,=\,-\,P^A
\end{equation}
 and equivalently for $Sp(2n)$ vectors  $P_ \alpha$:
 \begin{equation}
\IC_{\alpha\beta}\, P^\beta \,=\,P_ \alpha ; \qquad  \qquad
\IC^{\alpha\beta}\, P_ \beta \,=\,-\,P^\alpha
\end{equation}
\noindent
Since we have a product manifold the generic Killing vector
$k^i_{\Lambda}$ splits into
\begin{equation}\label{2kill}
  k^i_{\Lambda} \rightarrow {(k^i_{\Lambda},\, k^{\bar
  i}_{\Lambda});k^u_{\Lambda}}
\end{equation}
 and they can be determined in terms of the prepotentials of Special
 K\"{a}hler and Quaternionic geometry as follows:
\begin{eqnarray}\label{prep}
&&  k^i_{\Lambda}(z) \,=\,{\rm {i}} g^{ij^\star} \partial_{\bar
j}\,P_{\Lambda}(z,\, \bar z) \\
&&2 k^u_{\Lambda} \, \Omega ^x_{uv}\, =\,-
\nabla_v\,P^x_{\Lambda}(q)
\end{eqnarray}
where $\Omega ^x_{uv}$ are the $SU(2)$-valued components of the
quaternionic curvature strictly related to the three complex
structures existing on a quaternionic manifold. Note that as a
consequence of quaternionic geometry (see Appendix) the quaternionic
prepotentials satisfy the following "harmonic" equation: \footnote
{Here and in the following we have set  $\lambda\,=\,-1$ where  $\lambda$ is
the scale (defined in the Appendix) of ${\mathcal
M}^{(hyper)}$. This is required by four dimensional
supersymmetry of the Lagrangian (see \cite{dff},\cite{a1})}.

\begin{equation}\label{lapla}
\nabla_u\,\nabla^u P^x_{\Lambda}\,=\, -4n_H\, P^x_{\Lambda}
\end{equation}

For the purpose of making the paper self contained we report now
the Lagrangian and the transformation laws of the $N=2$ Lagrangian
as given in reference \cite{a1}.We limit ourselves to report the Lagrangian up
to 4-fermion terms and the supersymmetry transformation laws up to 3-fermion terms since this is
sufficient for our treatment. We have:
\vskip 0.5cm
{\it N=2 Supergravity lagrangian}

\begin{eqnarray} \label{lagn2}
({\rm det}V)^{-1}\, {\mathcal L} &=&
-\frac{1}{2} R +
g_{ij^\star}\nabla^\mu z^i \nabla_\mu \bar z^{j^\star}+
h_{uv}\nabla_\mu q^u \nabla^\mu q^v +
{{\epsilon^{\mu\nu\lambda\sigma}}\over{\sqrt{-g}}   }
\left( \bar\psi^A_\mu\gamma_\sigma \rho_{A\nu\lambda}
-  \bar\psi_{A\mu} \gamma_\sigma \rho^A_{\nu\lambda} \right )
\nonumber \\
&-& {{\rm i}\over2}g_{ij^\star} \left(\bar\lambda^{iA}\gamma^\mu
\nabla_\mu\lambda^{j^\star}_A
+\bar\lambda^{j^\star}_A \gamma^\mu \nabla_\mu\lambda^{iA}\right )
-{\rm i}\left (\bar\zeta^\alpha\gamma^\mu\nabla_\mu\zeta_\alpha
+\bar\zeta_\alpha\gamma^\mu \nabla_\mu \zeta^\alpha \right) \nonumber \\
&+& {\rm i}\left(
\bar {\cal N}_{\Lambda\Sigma}{\cal F}^{-\Lambda}_{\mu\nu}{\cal F}^{-\Sigma
\mu\nu} -
{\cal N}_{\Lambda\Sigma} {\cal F}^{+\Lambda}_{\mu\nu}{\cal F}^{+ \Sigma
{\mu\nu}}\right )+ \Big\{ -g_{ij^\star}
 \nabla_\mu \bar z^{j^\star} \bar \psi^\mu_A \lambda^{i A} \nonumber\\
&-& 2 {\cal U}^{A\alpha}_u \nabla_\mu q^u
\bar \psi_A^\mu \zeta _\alpha
+ g_{ij^\star}  \nabla _\mu \bar z^{j^\star}
\bar \lambda^{iA} \gamma^{\mu\nu} \psi_{A\nu}
+ 2{\cal U}^{\alpha A}_u \nabla_\mu q^u
\bar \zeta_\alpha \gamma^{\mu\nu}
\psi_{A\nu}+{\rm h.c.}\Big\} \nonumber \\
&+& \{
{\cal F}^{-\Lambda}_{\mu\nu}
{\rm Im }\,{\cal N}_{\Lambda\Sigma}\,
{\lbrack} 4 L^\Sigma  \bar \psi^{A\mu}
\psi^{B\nu} \epsilon_{AB}-4{\rm i}
{\bar f}^\Sigma_{i^\star}\bar \lambda^{i^\star}_A \gamma^\nu
\psi_B^\mu \epsilon^{AB}  \nonumber \\
&+& \frac{1}{2}
\nabla_i f^\Sigma_j
\bar \lambda^{iA} \gamma^{\mu\nu} \lambda^{jB}\epsilon_{AB}-
L^\Sigma \bar\zeta_\alpha\gamma^{\mu\nu} \zeta_\beta
C^{\alpha\beta}
{\rbrack}+{\rm h.c.}\}
\nonumber  \\
&+& \bigl[2S_{AB}\bar\psi^A_\mu\gamma^{\mu\nu}\psi^B_\nu +
{\rm i} g_{ij^\star} W^{iAB} \bar\lambda^{j^\star}_A\gamma_\mu \psi_B^\mu+
 2{\rm i} N^A_\alpha\bar\zeta^\alpha\gamma_\mu \psi_A^\mu \nonumber \\
&+&
{\cal M}^{\alpha\beta}{\bar \zeta}_\alpha
\zeta_\beta +{\cal M}^{\alpha}_{\phantom{\alpha}iB}
{\bar\zeta}_\alpha \lambda^{iB} + {\cal M}_{ij\,AB}
{\bar \lambda}^{iA} \lambda^{j
B} + \mbox{h.c.}\bigr]
-{\mathcal V}\bigl ( z, {\bar z}, q \bigr ).
\end{eqnarray}
where we have set  ${\mathcal F}^{\pm \Lambda}_{\mu\nu}=\frac {1}{2} ({\mathcal F}^\Lambda_{\mu\nu}\pm
{{\rm i}\over2}\epsilon^{\mu\nu\rho\sigma} {\mathcal F}_{\rho\sigma}^\Lambda)$, ${\mathcal F}^\Lambda_{\mu\nu}$
being the field-strengths of the vectors $A^\Lambda_\mu$.
Furthermore $L^\Lambda(z,\,\bar z)$ are the covariantly holomorphic sections of
the Special Geometry, $f^\Lambda_{i}\equiv \nabla_i L^\Lambda$ and the kinetic matrix
${\cal N}_{\Lambda\Sigma}$ is constructed in terms of $ L^\Lambda$ and its magnetic dual
according to reference \cite{a1}.
The normalization of the kinetic term for the quaternions
depends on the scale $\lambda$ of the quaternionic manifold
 for which we have chosen the value $\lambda=-1$,(see footnote 3).
Finally the  mass matrices of the spin $ \frac {1}{2}$ fermions  ${\mathcal M}^{\alpha \,\beta}$, ${\mathcal
 M}_{AB \,ij}$, ${\mathcal M}^{\alpha}_{iA}$ (and
 their hermitian conjugates) and the scalar potential $\mathcal V$ they are given by\footnote{There are misprints in the
 equation (\ref{pesamatrice}) as given in reference  \cite{a1} which have been corrected.}:
\begin{eqnarray}\label{mass2}
{\cal M}^{\alpha\beta}  &=&- \, {\cal U}^{\alpha A}_u \, {\cal
U}^{\beta B}_v \, \varepsilon_{AB}
\, \nabla^{[u}   k^{v]}_{\Lambda}  \, L^{\Lambda} \\
{\cal M}^{\alpha }_{\phantom{\alpha} iB} &=& -4 \, {\cal
U}^{\alpha}_{B  u} \, k^u_{\Lambda} \,
 f^{\Lambda}_i \\
{\cal M}_{AB \,\, ik} &=&  \,  \epsilon_{AB} \,
g_{l^{\star} [i}
 f_{k]}^\Lambda  k^{l^{\star}}_ \Lambda \,+ \frac {1}{2}
{\rm {i}}  P_ {\Lambda AB} \,
\nabla_i f^\Lambda _k  \label{pesamatrice}
\end{eqnarray}

\begin{equation}\label{potn2}
 {\mathcal V}(z,\bar z, q)\,=\,g^2 \Bigl[\left(g_{ij^\star} k^i_\Lambda k^{j^\star}_\Sigma+4 h_{uv}
k^u_\Lambda k^v_\Sigma\right) \bar L^\Lambda L^\Sigma
%\nonumber\\
+ g^{ij^\star} f^\Lambda_i f^\Sigma_{j^\star}
{\cal P}^x_\Lambda{\cal P}^x_\Sigma
-3\bar L^\Lambda L^\Sigma{\cal P}^x_\Lambda
{\cal P}^x_\Sigma\Bigr]\ .
\end{equation}
The supersymmetry transformation laws leaving invariant (\ref{lagn2}) are:
\vskip 0.5cm
{\it Supergravity transformation rules of the (left--handed) Fermi  fields}:

\begin{eqnarray}
\delta\,\psi _{A \mu} &=& {\cal D}_{\mu}\,\epsilon _A\,
 + \left ( {\rm i} \, g \,S_{AB}\eta _{\mu \nu}+
\epsilon_{AB} T^-_{\mu \nu} \right) \gamma^{\nu}\epsilon^B
 \label{trasfgrav} \\
\delta \,\lambda^{iA}&=&
 {\rm i}\,\nabla _ {\mu}\, z^i
\gamma^{\mu} \epsilon^A
+G^{-i}_{\mu \nu} \gamma^{\mu \nu} \epsilon _B \epsilon^{AB}\,+\,
W^{iAB}\epsilon _B
\label{gaugintrasfm}\\
 \delta\,\zeta _{\alpha}&=&{\rm i}\,
{\cal U}^{B \beta}_{u}\, \nabla _{\mu}\,q^u
\,\gamma^{\mu} \epsilon^A
\epsilon _{AB}\,C_{\alpha  \beta}
\,+\,N_{\alpha}^A\,\epsilon _A \label{iperintrasf}
\end{eqnarray}
where $T^-_{\mu\nu} = 2{\rm i}
{\ Im} {\cal N}_{\Lambda\Sigma} L^{\Sigma}
F_{\mu\nu}^{\Lambda -}$ and $G^{i-}_{\mu\nu} = - g^{i{{j}^\star}}
 \bar f^\Gamma_{{j}^\star}
{\ Im } {\cal N}_{\Gamma\Lambda}
  {F}^{\Lambda -}_{\mu\nu}$.
\vskip 0.5cm
{\it Supergravity transformation rules of the Bose  fields}:

\begin{eqnarray}
\delta\,V^a_{\mu}&=& -{\rm i}\,\bar {\psi}_{A
\mu}\,\gamma^a\,\epsilon^A -{\rm i}\,\bar {\psi}^A _
\mu\,\gamma^a\,\epsilon_A\\
\delta \,A^\Lambda _{\mu}&=&
2 \bar L^\Lambda \bar \psi _{A\mu} \epsilon _B
\epsilon^{AB}\,+\,2L^\Lambda\bar\psi^A_{\mu}\epsilon^B \epsilon
_{AB}\nonumber\\
&+&{\rm i} \,f^{\Lambda}_i \,\bar {\lambda}^{iA}
\gamma _{\mu} \epsilon^B \,\epsilon _{AB} +{\rm i} \,
{\bar f}^{\Lambda}_{{i}^\star} \,\bar\lambda^{{i}^\star}_A
\gamma _{\mu} \epsilon_B \,\epsilon^{AB} \label{gaugtrasf}\\
\delta\,z^i &=& \bar{\lambda}^{iA}\epsilon _A \label{ztrasf}\\
\delta\,z^{{i}^\star}&=& \bar{\lambda}^{{i}^\star}_A \epsilon^A
\label{ztrasfb}\\
  \delta\,q^u &=& {\cal U}^u_{\alpha A} \left(\bar {\zeta}^{\alpha}
  \epsilon^A + C^{\alpha  \beta}\epsilon^{AB}\bar {\zeta}_{\beta}
  \epsilon _B \right)
 \end{eqnarray}
The gauge shifts for the three kinds of fermions, gravitinos $\psi_{A\mu}$,(A=1,2),
gauginos $\lambda^{iA}\,( i=1,\dots n_V)$ and hyperinos $
\zeta^{\alpha},(\alpha =1,\dots,n_H)$ appearing both in
the Lagrangian and in the supersymmetry transformation laws are given by  (here and in the following
we set $g=1$ for the gauge coupling constant appearing in the Lagrangian
and in the transformation laws)\cite{a1}:

\begin{eqnarray}\label{trapsi2}
S_{AB}&=& {\rm i} \frac {1}{2} P_{AB\, \Lambda} \,
 L^{\Lambda} \equiv  {\rm i} \frac {1}{2} P_{AB}
 \equiv {\rm i} \frac {1}{2} P^x \sigma^x_{AB} \\
\label{tralam} W^{i\,AB}&=& {\rm {i}} P^{AB}_{ \Lambda}\,g^{ij^\star}
f^{\Lambda}_{j^\star} + \epsilon^{AB} k^i_{\Lambda}{\overline
L}^{\Lambda}\equiv  {\rm {i}} \nabla^i P^{AB} + \epsilon^{AB}
k^i\\
N^A_{\alpha}&=& 2\,{\mathcal U}^A_{\alpha u} \,k^u_{\Lambda}
{\overline L}^{\Lambda}\equiv 2\,{\mathcal U}^A_{\alpha u} \,\overline{k}^u\\
N_A^{\alpha}&=& -2\,{\mathcal U}^{\alpha }_{Au} \,k^u_{\Lambda} L^{\Lambda}
\equiv -2\,{\mathcal U}^{\alpha }_{Au} \,k^u_\\
\end{eqnarray}
where \footnote{ We use Pauli matrices with
 both indices in the upper or lower position so that they are symmetric.Note
 that $ (\sigma_{AB}^x)^\star \,=\,- \sigma^{x \,AB} $.}:
\begin{eqnarray}\label{deff}
P_{AB \Lambda} L^\Lambda \,&=& \, P^x_{\Lambda} L^\Lambda \,
 \sigma_{AB}^x \equiv  P^x \,
 \sigma_{AB}^x,\,\,\, (x=1,2,3)\\
 (P_{\Lambda AB} L^\Lambda)^\star \,&=&\,-P_{\Lambda}^{AB} \bar L^\Lambda \,=\
 -P_\Lambda^x \bar L^\Lambda \sigma^{xAB}\equiv - P^x \sigma^{xAB}
\end{eqnarray}
and we have further defined
\begin{eqnarray}\label{uffa}
k^i_{\Lambda} \bar L^{\Lambda}\, &=&\,k^i \\
\label{riuffa}k^u_{\Lambda} \bar L^{\Lambda}&=& \overline{k}^u \\
\label{riuffa2}k^u_{\Lambda}  L^{\Lambda}&=& k^u \\
 P_{AB \Lambda} \, L^{\Lambda} & =&  P_{AB}\\
 P^{AB}_{ \Lambda}\,g^{ij^\star}
f^{\Lambda}_{j^\star} &=&
\nabla_{j^\star} \bar L^{\Lambda}P^{AB}_{
\Lambda}\,g^{ij^\star}=\nabla^i P^{AB}
\end{eqnarray}
Taking into account the definitions (\ref{uffa}), (\ref{riuffa})
we see that  the $\delta {\lambda} ^{iA}$ and $\delta {\chi_\alpha}$
shifts are all covariant derivatives of the quaternionic and K\"ahler prepotentials
\footnote{Note that $\nabla^i P_\Lambda {\bar L}^\Lambda$ cannot be written as a total derivative
since $P_\Lambda {\bar L}^\Lambda \,=\,0$, so no gauge invariant prepotential exists
for vector multiplets as in the $N=1$ case (see (\ref{dlambda})) . This should no come as a surprise since a prepotential
having the interpretation of "superpotential" should be related to
the gravitino mass and it is $S_{AB}$ (quaternionic prepotential)  and not the
Hodge -K\"ahler prepotential $P_\Lambda$ which enters in it.
Indeed no gauge invariant scalar exists for vector multiplets that could enter
 in the spin $\frac {3}{2} $ mass term.}:

\begin{eqnarray}\label{new}
W^{i\,AB}&=& {\rm {i}} \nabla^i P^{AB}\,+\,\epsilon^{AB} \nabla^i P_\Lambda {\bar L}^\Lambda  \\
N^A_{\alpha}&=& - \frac {1}{3} {\mathcal U}^A_{\alpha u} \, \Omega^{x uv} \,
 \nabla_v  P^x\,=\ 2\,{\mathcal U}^A_{\alpha u} \,\overline{k}^u\\
N^{\alpha}_A&=& =\- 2\,{\mathcal U}^{\alpha }_{Au} \,k^u\\
\end{eqnarray}

 The gradient flow equations (\ref {gfN})and (\ref {cast})
 adapted to the present case are:
\begin{eqnarray}\label{gradflo2}
\nabla_k \,W^{i\,AB} &=& 2\, \delta^i_k \,S^{AB} -
\epsilon^{AB}\,g^{i
j^\star}\,k_{k \,\Lambda}\,f^{\Lambda}_{j^\star} \\
\nabla_{k^\star} \,W^{i \,AB} &=& - g^{ij^\star}  {\mathcal
 M}^{AB}_{k^\star j^\star}   + \,\frac {1}{2} \epsilon^{AB} g^{ij^\star}
 \left ( f^\Lambda_{j^\star} k_{\Lambda k^\star}
 \,+\,f^\Lambda_{k^\star} k_{\Lambda j^\star} \right) \\
\nabla_u \, W_{AB}^{j^\star}&=& - \frac {1}{2} g^{ij^\star} \,{\mathcal
M}^{\alpha}_{i(B} \,{\mathcal U}_{A)\alpha u} \\
\nabla_i
N^{\alpha}_A&=& \frac {1} {2}\,{\mathcal
M}^{\alpha}_{iA} \\
\label{dn} {\mathcal U}^{u\,B \alpha}\,\nabla_u\,N^{A \beta}&=& 4\,C^{\alpha
\beta}\, S^{AB}\,+\,\epsilon^{AB}\,{\mathcal M}^{\alpha \,\beta} \\
\label{du} \nabla_u S_{AB}&=& - \frac {1}{2}{\mathcal U}_{\alpha u
(A}\,N_{B)}^{\alpha} \\
\label{di} \nabla_i S_{AB}&=& \frac {1}{2} \, W_{(AB)}^{j^\star}
\, g_{i
j^\star} \\
\label{dib} \nabla_{i^\star}\,S_{AB}&=&0
\end{eqnarray}
where we have set $W_{AB}^{j^\star} \equiv (W^{j\,AB})^\star$ and ${\mathcal
 M}^{AB}_{k^\star j^\star} \equiv ({\cal M}_{AB \,\, kj})^\star $.
 \par
 Note that in the manipulations performed from (\ref{new}) to (\ref{gradflo2}) one has to use the Gauged Special
 Geometry identities \cite{dff}:
\begin{eqnarray}\label{orto}
  P_\Lambda L^\Lambda\,&=&\, P_\Lambda {\bar L}^\Lambda\,=\,0 \\
  k^i_\Lambda L^\Lambda\,&=&\,k^{i\star}_\Lambda {\bar L}^\Lambda\,=\,0
\end{eqnarray}

We also note that ${\mathcal M}^{\alpha \,\beta}$ can be written in terms
of the (traceless part of the)
anticommutator of two
covariant derivatives
\footnote{Note that in eq. (\ref{anti}) the symmetric part of the anticommutator is automatically
traceless because ${\mathcal U}^{u(A}_\alpha {\mathcal U}^{B)v}_\beta h_{uv}=0$}:
\begin{equation}\label{anti}
{\cal M}_{\alpha\beta} \,=\, {\rm {i}} \frac {1}{6} {\cal U}^ {u}_{A(\alpha} {\cal
U}^{v}_{\beta)B} \nabla_{(u} \nabla_{v)} P^{AB}
\end{equation}
This is a consequence of the basic relations of gauged quaternionic geometry
given in Appendix. Equation (\ref{anti}) is analogous to
what happened for chiral multiplet in the $N=1$ case, see equation
(\ref{def4}). We also note that equation (\ref{dn}) is equivalent to equation (\ref{madre})
of the Appendix where we put $\lambda\,=\,-1$.

Using the identity (\ref{pot}) adapted to the present case:
\begin{equation} \label{ward}
\delta_B^A {\overline V} \,= \,-12(S^{MA})^\star \, S_{MB}\,+\,g_{ij^\star}W^{iMA}\,W_{MB}^{j^\star}\,
+\,2N^A_\alpha \, N^\alpha_B
\end{equation}
 one
may compute the scalar potential which turns out to be the one given in eq.
(\ref{potn2})\cite{dff}, \cite{a1}.
Using equations (\ref{du}), (\ref{di}),
(\ref{dib}) the scalar potential (\ref{potn2}) can be also rewritten in the
following way:
\begin{equation} \label{pot2}
{\mathcal V}\,=-\,6\,S_{AB}\, (S^{AB})^\star\,+\,2 g^{ij^\star} \nabla_i
S_{AB}\,\nabla_{j^\star} (S^{AB})^\star \,+\,4\,\nabla_u
S_{AB}\,\nabla^{u}(S^{AB})^\star\,+\,g_{ij^\star}k^i k^{j^\star}
\end{equation}
\noindent
 Finally we note that the last term in equations (\ref{new}),(\ref{tralam}) has a similar structure as
the $N=2$ central charge \cite{cdfv}, but the $SU(2)$ valued prepotential
 $ P^{AB}$ adds a symmetric part to
$W^{i\,AB} $ .
However unbroken $N=2$ supersymmetry is still controlled by a gradient flow equation
\cite{bhlt}, \cite{gs}, \cite{bgs} which is equivalent to $k^i\,=\,0$ and
to extremize in the moduli space $P^x$.\footnote{These conditions are generally
too restrictive if only $N=1$ supersymmetry is preserved. For instance, in the
example of ref. \cite{fgp} $N^A_\alpha$ does not vanish.}
We note in particular that if $P^x \,=\,0$ the supersymmetry flow has always vanishing
 potential (no $AdS$ vacua).
On the other hand if $k^i\,=\,0$ the supersymmetry flow is controlled by the "superpotential"
$P^x \, P^x$ whose extrema in the full moduli space (at $P^x \neq 0$) imply
$\delta {\lambda} ^{iA}$\,=\, $\delta {\zeta_\alpha}\,=\,0 $.
For abelian gauging $k^i= k^i_\Lambda\,=0$ and in absence of hypermultiplets , for constant $P^x_\Lambda$,
we retrieve the situation discussed in the literarature
\cite{cdfgkdv}, \cite{ckrrs}, \cite{bhlt}, \cite{bgs}.
\par
We note that he supersymmetric flow of the hypermultiplets ( at points where the scalars have vanishing velocity)
implies a vanishing value of the Killing vectors:  $k^u_\Lambda L^\Lambda \,=\,0$.
 Since the covariant holomorphic section $L^\Lambda = L^\Lambda(z,\bar z)$ is complex,
 this implies that the values $ q^u=q^u_0$ for which $k^u_\Lambda L^\Lambda \,=\,0$
 are the fixed points \footnote{The same
 is true for the Special K\"ahler manifolds
 in the case of non Abelian isometries.}of the group generated by the two (real) Lie Algebra elements
 $ k^u_\Lambda {\rm Im}L^\Lambda, k^u_\Lambda {\rm  Re}
L^\Lambda$. This group depends on ${\rm Im}L^\Lambda$
 and ${\rm Re}L^\Lambda$. If
\begin{equation}\label{group}
f^{\Delta}_{\phantom{\Delta}\Lambda\Sigma}\,{\rm
Im}L^ \Lambda \,{\rm Re}L^\Sigma \,=\,0
\end{equation}
the two-dimensional gauge group is Abelian. If either ${\rm Im}L^\Lambda$ or ${\rm Re}L^\Lambda$ is zero
then we have a one-dimensional gauge group; otherwise it may be any subgroup generated by the two elements.
Of course for abelian isometries $f^{\Delta}_{\phantom{\Delta}\Lambda\Sigma}\,=\,0$ and the group is always
a two-dimensional subgroup of the gauge group.
It  is interesting to observe that in five dimensions the corresponding section $L^\Lambda$ is real
so that the group generated by $k^u_\Lambda L^\Lambda$ is always a one-dimensional
subgroup of the isometry group.
Furthemore we note that the condition $k^u_\Lambda L^\Lambda\,=\,0$,
 taking into account the equation (\ref{quatide}),implies the following consistency
 condition on the quaternionic prepotentials:
\begin{equation}\label{cons}
-  {\lambda} \, \varepsilon^{xyz} \,P_\Lambda^y  \, P_\Sigma^z\,L^\Lambda \bar L^\Sigma= \,
f^{\Delta}_{\phantom{\Delta}\Lambda\Sigma}\, P_\Delta^{x}\,L^\Lambda \bar L^\Sigma
\end{equation}
which in the Abelian case reduces to
\begin{equation}\label{para}
\varepsilon^{xyz} \,
 P_\Lambda^y  \, P_\Sigma^z\,\,L^\Lambda \bar L^\Sigma= \,0.
\end{equation}
Defining $P^x(L)\,=\,P^x_\Lambda L^\Lambda$ (and setting $\lambda=-1$), equation (\ref{cons})
can be rewritten in the suggestive form:
\begin{equation}\label{sugg}
\overrightarrow{P}(L)\,\wedge\,\overrightarrow{P}( \bar L)\,=\,\overrightarrow{P}(L\times \bar L)
\end{equation}
where
\begin{equation}\label{defff}
L\times \bar L \equiv f^{\Delta}_{\phantom{\Delta}\Lambda\Sigma} L^\Lambda \bar L^\Sigma
\end{equation}

\section{Dual Quaternionic Manifolds and the gauging of their isometries.}

A particular interesting case where some "universal gauging" can be studied in a fairly general way
is the special situation when the hypermultiplet manifold of quaternionic dimension $n+1$ is obtained
by $c$--map \cite{cgf} from a Special K\"ahler manifold of complex dimensions $n$.

These manifolds, called dual quaternionic manifolds in reference \cite{cgf}, have a "solvable group
of motion" whose Solvable Lie Algebra has dimension $2n+4$ \cite{adfft}. This solvable Lie Algebra is associated to the rank
one coset $SU(1,n+2)/SU(n+2)\otimes U(1)$ and contains, as particular case, the "universal hypermultiplet"
parametrizing $SU(1,2)/U(2)$ \cite{cgf}. These symmetries are always present even if the Special
K\"ahler manifold has no isometries at all.

In Calabi--Yau compactifications for Type IIB strings down to $D=4$, $n=h_{1,1}$, and
the solvable algebra of rank one is related to $2h_{1,1}+2$ shift-- symmetries of the $RR$--scalars, one
shift symmetry of the $NS$--axion (dual to $b_{\mu\nu}$) making the $2h_{1,1}+3$ nilpotent part of the group.
The remaining Cartan generator is related to the scale symmetry of the dilaton. The maximal abelian
ideal has dimension $h_{1,1}+2$ (of which $h_{1,1}+1$ are $RR$ abelian shifts).

In the case of the universal hypermultiplet the nilpotent subalgebra is three--dimensional,
 it is the Heisenberg algebra considerd in reference \cite{bb}, where also the discrete remnant, after brane instanton
 corrections to Hypergeometry, was considered.

For a "dual quaternionic manifold" one can then always gauge the solvable group (non--abelian
gauging) or restrict to the abelian gauging of its "maximal abelian ideal" of dimension $n+2$.
To achieve this gauging one must at least have $2n+3$ vector multiplets ($n+1$ in the abelian case).
\footnote{ For Calabi--Yau compactifications this would imply that $h_{2,1}-h_{1,1} =\frac{1}{2}\chi\geq 1$
in Type IIB ($\chi \rightarrow - \chi$ in Type IIA).}
\par
It is interesting that all the existing examples of gauging are particular cases
of this general framework.\par
The gauging of the two shift--symmetries of the universal multiplet was considered in references
\cite{ps}, \cite{m}, \cite{tava} and correspond to the $n=0$ case; it is obtained by turning on
the $H$--fluxes of the two field--strengths of the $NS$ and $RR$ two--forms on a Calabi--Yau threefold.
This case requires $h_{2,1}\geq 2$.
\par
Another case considered in the literature is the case when the maximal compact subgroup of
the isometry group is gauged. In order this to be the case we
 may consider dual quaternionic spaces
which are coset spaces, in which case also the Special K\"ahler manifold is a coset. The most general
abelian compact gauging is obtained by gauging the Cartan subalgebra of the maximal compact subgroup.
For the unitary series this has dimension $n+2$ and for $n=0$ reduces to the gauging of $U(1)^2$
of the universal multiplet considered in reference \cite{cdkv}.
\par
 For the $G_2/SO(4)$ manifold dual to
$SU(1,1)/U(1)$ special K\"ahler manifold, the gauging of the isometries was considered in reference \cite{Betal}.

\appendix
\section*{Appendix: Glossary of Quaternionic Geometry}
\setcounter{equation}{0}
\addtocounter{section}{1}

Both a quaternionic or a HyperK\"ahler manifold ${\cal M}^{(hyper)}$
are  $4 n$-dimensional real manifolds
endowed with a metric $h$:
\begin{equation}
d s^2 = h_{u v} (q) d q^u \otimes d q^v   \quad ; \quad u,v=1,\dots,
4 n_H
\label{qmetrica}
\end{equation}
and three complex structures
\begin{equation}
(J^x) \,:~~ T({\cal HM}) \, \longrightarrow \, T({\cal HM}) \qquad
\quad
(x=1,2,3)
\end{equation}
that satisfy the quaternionic algebra
\begin{equation}
J^x J^y = - \delta^{xy} \, \bfone \,  +  \, \epsilon^{xyz} J^z
\label{quatalgebra}
\end{equation}
and respect to which the metric is hermitian:
\begin{equation}
\forall   \mbox{\bf X} ,\mbox{\bf Y}  \in   T{\cal HM}   \,: \quad
h \left( J^x \mbox{\bf X}, J^x \mbox{\bf Y} \right )   =
h \left( \mbox{\bf X}, \mbox{\bf Y} \right ) \quad \quad
  (x=1,2,3)
\label{hermit}
\end{equation}
From eq.s (\ref{quatalgebra}), (\ref{hermit}) it follows that one can introduce
a triplet of  2-forms
\begin{equation}
\begin{array}{ccccccc}
K^x& = &K^x_{u v} d q^u \wedge d q^v & ; &
K^x_{uv} &=&   h_{uw} (J^x)^w_v \cr
\end{array}
\label{iperforme}
\end{equation}
that provides the generalization of the concept of K\"ahler form
occurring in  the complex case. The triplet $K^x$ is named
the {\it HyperK\"ahler} form. It is an $SU(2)$ Lie--algebra valued
2--form  in the same way as the K\"ahler form is a $U(1)$ Lie--algebra
valued 2--form. In $N=1$ supersymmetry there is a single complex structure
and the scalar manifold has a K\"ahler structure implying
that the K\"ahler 2--form is closed. If  supersymmetry is local
the K\"ahler 2--form can be identified with the curvature of
the $U(1)$ line--bundle and in this case the manifold is called a
 Hodge--K\"ahler manifold, while for rigid supersymmetry the line bundle is flat.
 Similar steps can be also taken here and lead to two possibilities:
either HyperK\"ahler or Quaternionic manifolds.
\par
Let us  introduce a principal $SU(2)$--bundle over ${\cal M}^{(hyper)}$
. Let $\omega^x$ denote a
connection on such a bundle.
To obtain either a HyperK\"ahler or a Quaternionic manifold
we must impose the condition that the HyperK\"ahler 2--form $K^x$
is covariantly closed with respect to the connection $\omega^x$:
\begin{equation}
\nabla K^x \equiv d K^x + \epsilon^{x y z} \omega^y \wedge
K^z    \, = \, 0
\label{closkform}
\end{equation}
The only difference between the two kinds of geometries resides
in the structure of the ${SU(2)}$--bundle.

A HyperK\"ahler manifold is a $4 n$--dimensional manifold with
the structure described above and such that the ${SU(2)}$--bundle
is flat.

 Defining the ${SU(2)}$--curvature by:
\begin{equation}
\Omega^x \, \equiv \, d \omega^x +
{1\over 2} \epsilon^{x y z} \omega^y \wedge \omega^z
\label{su2curv}
\end{equation}
in the HyperK\"ahler case we have:
\begin{equation}
\Omega^x \, = \, 0
\label{piattello}
\end{equation}
Viceversa
a Quaternionic manifold is a $4 n$--dimensional manifold with
the structure described above and such that the curvature
of the ${SU(2)}$--bundle
is proportional to the HyperK\"ahler 2--form.
Hence, in the quaternionic case we can write:
\begin{equation}
\Omega^x \, = \, { {\lambda}}\, K^x
\label{piegatello}
\end{equation}
where $\lambda$ is a non vanishing real number which, as we shall see,
 sets the scale of the manifold ${\cal M}^{(hyper)}$.
\par
As a consequence of the above structure the Quaternionic manifold  has a
holonomy group of the following type:
\begin{eqnarray}
{\rm Hol}({\cal M}^{(hyper)})&=& SU(2)\otimes {\cal H} \quad (\mbox{quaternionic})
\nonumber \\
{\cal H} & \subset & Sp (2n_H,\IR)
\label{olonomia}
\end{eqnarray}
Introducing flat
indices $\{A,B,C= 1,2\}\, \{\alpha,\beta,\gamma = 1,.., 2n\}$  that run,
respectively, in the fundamental representations of $SU(2)$ and
$Sp(2m,\IR)$, we can find a vielbein 1-form
\begin{equation}
{\cal U}^{A\alpha} = {\cal U}^{A\alpha}_u (q) d q^u
\label{quatvielbein}
\end{equation}
such that
\begin{equation}
\label{metric} h_{uv} = {\cal U}^{A\alpha}_u {\cal U}^{B\beta}_v
\IC_{\alpha\beta}\epsilon_{AB}
\label{quatmet}
\end{equation}
where $C_{\alpha \beta} = - C_{\beta \alpha}$ and
$\epsilon_{AB} = - \epsilon_{BA}$ are, respectively, the flat $Sp(2n)$
and $Sp(2) \sim SU(2)$ invariant metrics.
The vielbein ${\cal U}^{A\alpha}$ is covariantly closed with respect
to the $SU(2)$-connection $\omega^z$ and to some $Sp(2m,\IR)$-Lie Algebra
valued connection $\Delta^{\alpha\beta} = \Delta^{\beta \alpha}$:
\begin{eqnarray}
\nabla {\cal U}^{A\alpha}& \equiv & d{\cal U}^{A\alpha}
+{i\over 2} \omega^x  \sigma_x^{AB}
\wedge {\cal U}_B^{\alpha} \nonumber\\
&+& \Delta^{\alpha\beta} \wedge {\cal U}^{A\gamma} \IC_{\beta\gamma}
=0
\label{quattorsion}
\end{eqnarray}
\noindent
where $(\sigma^x)^{AB}\,=\,\epsilon^{AB}(\sigma^x)_A^{\phantom{A}C}$
and $(\sigma^x)_A^{\phantom{A}B}$ are the standard Pauli matrices.
Furthermore ${ \cal U}^{A\alpha}$ satisfies  the reality condition:
\begin{equation}
{\cal U}_{A\alpha} \equiv ({\cal U}^{A\alpha})^* = \epsilon_{AB}
\IC_{\alpha\beta} {\cal U}^{B\beta}
\label{quatreality}
\end{equation}
More specifically we can write a stronger
version of eq. (\ref{metric} \cite{bw}:
\begin{equation}
({\cal U}^{A\alpha}_u {\cal U}^{B\beta}_v + {\cal U}^{A\alpha}_v {\cal
 U}^{B\beta}_u) C_{\alpha\beta}= h_{uv} \epsilon^{AB}
\label{piuforte}
\end{equation}
\noindent
The inverse vielbein ${\cal U}^u_{A\alpha}$ is defined by
\begin{equation}
{\cal U}^u_{A\alpha} {\cal U}^{A\alpha}_v = \delta^u_v
\label{2.64}
\end{equation}
Flattening a pair of indices of the Riemann
tensor ${\cal R}^{uv}_{\phantom{uv}{ts}}$
we obtain
\begin{equation}
{\cal R}^{uv}_{\phantom{uv}{ts}} {\cal U}^{\alpha A}_u {\cal U}^{\beta B}_v =
-\,{{\rm i}\over 2} \Omega^x_{ts}
 (\sigma_x)^{AB} C^{\alpha \beta}+
 \IR^{\alpha\beta}_{ts}\epsilon^{AB}
\label{2.65}
\end{equation}
\noindent
where $\IR^{\alpha\beta}_{ts}$ is the field strength of the $Sp(2n_H)
$ connection:
\begin{equation}
d \Delta^{\alpha\beta} + \Delta^{\alpha \gamma} \wedge \Delta^{\delta \beta}
\IC_{\gamma \delta} \equiv \IR^{\alpha\beta} = \IR^{\alpha \beta}_{ts}
dq^t \wedge dq^s
\label{2.66}
\end{equation}
and
\begin{equation}\label{sympl}
\IR^{\alpha\beta}_{uv}=\frac{\lambda}{2}\epsilon_{AB}({\mathcal U}^{A\alpha}_u
{\mathcal U}^{B\beta}_v-{\mathcal U}^{A\alpha}_v {\mathcal U}^{B\beta}_u)+{\mathcal U}^{A\gamma}_u {\mathcal U}^{B\delta}_v
 \epsilon_{AB}
\IC^{\alpha\rho}\IC^{\beta\sigma} \Omega_{\gamma\delta\rho\sigma}\ ,
\end{equation}
where $\Omega_{\gamma\delta\rho\sigma}$ is a completely symmetric tensor.
The previous equations imply that the Quaternionic manifold is
an Einstein space with Ricci tensor given by \footnote {Our convention for
  the Riemann tensor are as follows: $ R^u_v \equiv d\Gamma^u_v +
  \Gamma^u_w \wedge \Gamma^w_v\,=\,R^u_{vrs}dq^r \wedge dq^s$
  where $\Gamma$
 is the Levi--Civita connection 1--form. Therefore the Ricci tensor is $R_{vs}= R^u_{vrs} \delta^r_u $}
\begin{equation}
\label{ricci}\cR_{uv}=\lambda (2+n_H) h_{uv}\ .
\end{equation}
Note that if the manifold is hyperK\"ahler, that is if equation (\ref{piegatello})
holds, then $\lambda=0$ and the manifold is Ricci flat.
\noindent
Eq. ~(\ref{sympl}) is the explicit statement that the Levi Civita connection
associated with the metric $h$ has a holonomy group contained in
$SU(2) \otimes Sp(2n_H)$. Consider now eq.s~(\ref{quatalgebra}),
(\ref{iperforme}) and~(\ref{piegatello}).
We easily deduce the following relation:
\begin{equation}
h^{st} K^x_{us} K^y_{tw} = -   \delta^{xy} h_{uw} +
  \epsilon^{xyz} K^z_{uw}
\label{universala}
\end{equation}
that holds true both in the HyperK\"ahler and in the Quaternionic case.
In the latter case, using eq. (\ref{piegatello}), equation
(\ref{universala}) can be rewritten as follows:
\begin{equation}
h^{st} \Omega^x_{us} \Omega^y_{tw} = - \lambda^2 \delta^{xy} h_{uw} +
\lambda \epsilon^{xyz} \Omega^z_{uw}
\label{2.67}
\end{equation}
\noindent
In the quaternionic case we can write:
\begin{equation}
\Omega^x_{A\alpha, B \beta} \equiv \Omega^x_{uv} {\cal U}^u_{A\alpha}
{\cal U}^v_{B\beta} =  i \lambda C_{\alpha\beta} (\sigma_x)_{AB}
\label{intrin}
\end{equation}
\noindent
Alternatively eq.(\ref{intrin}) can be rewritten in an intrinsic form as
\begin{equation}
\Omega^x =\,{\rm i}\, \lambda C_{\alpha\beta}
(\sigma _x)_{AB} {\cal U}^{\alpha A} \wedge {\cal
U}^{\beta B}
\label{su2cur}
\end{equation}
\noindent
whence we also get:
\begin{equation}
{i\over 2} \Omega^x (\sigma_x)^{AB} =
\lambda{\cal U}^{A\alpha} \wedge {\cal
U}^{B}_{\alpha}
\label{curform}
\end{equation}
There exist quaternionic manifolds which are homogeneous symmetric manifolds
 (a list of homogeneous symmetric quaternionic spaces are given in
\cite{a1}).

\noindent
In full analogy with the case of
K\"ahler manifolds, to each Killing vector
we can associate a triplet ${\cal
P}^x_\Lambda (q)$ of 0-form prepotentials.
Indeed we can set:
\begin{equation}
 {\bf i}_{ \bf k_\Lambda} \Omega^x =
- \nabla {\cal P}^x_\Lambda \equiv -(d {\cal
P}^x_\Lambda + \epsilon^{xyz} \omega^y {\cal P}^z_\Lambda)
\label{2.76}
\end{equation}
where $\nabla$ denotes the $SU(2)$-- covariant exterior derivative
 and ${\bf i}_{\bf k_\Lambda}$ denotes the contraction of a form with the vector $\bf k_\Lambda$.
 Using components the previous equation takes the form:
\begin{equation}\label{prepo}
2k^v_\Lambda \Omega_{uv}^x\,=\,\nabla_u P^x_\Lambda
\end{equation}

\par
Formula (\ref{prepo}) can be inverted to give the Killing vector in terms of the prepotential:
\begin{equation} \label{invert}
 h_{uw}k^w_\Lambda \, = \,-\frac {1}{6 \lambda^2} \Omega^x_{uv} \nabla^v P^x_\Lambda
\end{equation}
The three--holomorphic Poisson bracket is defined
as follows:

\begin{equation}
\{{\cal P}_\Lambda, {\cal P}_\Sigma\}^x \equiv {\bf i}_{\bf k_\Lambda}{\bf i}_{\bf k_\Sigma}K^x
  + \, \varepsilon^{xyz} \,
{\cal P}_\Lambda^y  \, {\cal P}_\Sigma^z
\label{quatpesce}
\end{equation}
where
\begin{equation}\label{defcon}
 \frac {1}{2} {\bf i}_{\bf k_\Lambda}{\bf i}_{\bf k_\Sigma}K^x
\equiv \lambda^{-1} \Omega^x_{uv} \, k^u_\Lambda \, k^v_\Sigma
\end{equation}
\noindent
and leads to the Poissonian realization of the Lie algebra
\begin{equation}
\left \{ {\cal P}_\Lambda, {\cal P}_\Sigma \right \}^x \, = \,
f^{\Delta}_{\phantom{\Delta}\Lambda\Sigma} \, {\cal P}_\Delta^{x}
\label{quatpescespada}
\end{equation}
which in components reads:
\begin{equation}
\lambda^{-1} \Omega^x_{uv} \, k^u_\Lambda \, k^v_\Sigma \, + \,
 \frac {1}{2} \, \varepsilon^{xyz} \,
{\cal P}_\Lambda^y  \, {\cal P}_\Sigma^z\,= \,  \frac {1}{2} \,
f^{\Delta}_{\phantom{\Delta}\Lambda\Sigma}\, {\cal P}_\Delta^{x}
\label{quatide}
\end{equation}
\noindent Defining for brevity as in (\ref{uffa}) $ k_u=k^u_\Lambda
L^\Lambda$, from the Killing equation:
\begin{equation}\label{kill1}
\nabla_u k_v\,+\,\nabla_v k_u \,=\,0,
\end{equation}
using
\begin{equation}\label{rule1}
\left[\nabla_u,\nabla_v \right]k_{w\Lambda}\,=\,-2\cR_{uvw}^{\phantom{uvw}l}k_{l\Lambda}
\end{equation}
 and the value
of the Ricci tensor (\ref{ricci}), one easily finds that $k_u$ is an
eigenfunction of the (covariant) Laplacian:
\begin{equation}\label{laplk}
\nabla_v \nabla^v k_u \,+ \, 2\lambda (2+n_H) k_u \,=\,0
\end{equation}
\noindent
Furthermore by double differentiation of $ {P}^{AB}_\Lambda $ ,using the identity :
\begin{equation}\label{uu}
{\mathcal U}^A_{\alpha u}{\mathcal U}^{B \alpha}_v\,=\,\frac {-\rm {i}}{2\lambda}\Omega^{AB}_{uv}\,-\,\frac {1}{2}
\epsilon^{AB}h_{uv}
\end{equation}
we  find that also the prepotential is an eigenfunction of the covariant Laplacian:
\begin{equation}\label{laplp}
\nabla_v \nabla^v P^x_\Lambda \,+\ 4n_H \lambda P^x_\Lambda \,=\,0
\end{equation}
As a check, inserting equation (\ref{invert}) into (\ref{laplk}) and commuting the covariant derivative
 with the Laplacian using the rule:
\begin{equation}\label{rule2}
 \left[\nabla_u,\nabla_v \right]P^x\,=\,2\epsilon^{xyz}\Omega^y_{uv}P^z
\end{equation}
we find that (\ref{laplk}) and (\ref{laplp}) are indeed consistent.
Using equation (\ref{laplp}) and (\ref{prepo}) we also find:
\begin{equation}
    P^x_\Lambda=-\frac{1}{2n_H\,\lambda}\nabla^u\,k_\Lambda^v\,\Omega^x_{uv}
\end{equation}

 \noindent Finally we note that since $\nabla_u k_v \equiv
\nabla_{[u} k_{v]}$ is a 2--form we can expand it on a basis of
2--form given by $\Omega^{AB}_{uv}$ and \,$ {\cal U}^{\alpha A}_
{[u} \, {\cal U}^{\beta B}_{v]} \, \epsilon_{AB}$ which is part of
the symplectic curvature 2--form given in equation (\ref{sympl}).

\noindent
Indeed one can write:
\begin{equation}\label{madre}
\nabla_u k_v\,=\, \frac {1}{2\lambda}\Omega^{x}_{uv}P^x\,-\,\frac {1}{2}{\cal U}^{\alpha A}_ {[u} \, {\cal
U}^{\beta B}_{v]} \, \epsilon_{AB} {\mathcal M}_{\alpha \beta}
\end{equation}
where ${\mathcal M}_{\alpha \beta}$ is the hyperino mass matrix defined as
the complex conjugate of eq.(\ref{mass2}).
Equation (\ref{madre}) can be easily proved to hold by inverting the 2--form ${\cal U}^{\alpha A}_ {[u} \, {\cal
U}^{\beta B}_{v]} \, \epsilon_{AB}$ on the r.h.s. of (\ref{madre}) and antisymmetrizing
 either in the $SU(2)$ indices or in the
symplectic indices.

\section*{Acknowledgements}
We heartly thank A.Ceresole for enlightening discussions
 and interest in our paper.We also acknowledge an important discussion with
 R. Stora.The work of R. D. has been supported
in part by the European Commission RTN network HPRN-CT-2000-00131
(Politecnico di Torino). The work of S. F. has been supported in
part by the European Commission RTN network HPRN-CT-2000-00131,
(Laboratori Nazionali di Frascati, INFN) and by the D.O.E. grant
DE-FG03-91ER40662, Task C.


\begin{thebibliography}{90}
\bibitem{maiafe} S. Ferrara and L. Maiani, {\it { An Introduction to supersymmetry
breaking in extended Supergravity}}, Proceedings of SILARG
V,Bariloche, Argentina,(1985),O.Bressan, M.Castagnino and
V.Hamity editors, World Scientific (1985)
\bibitem{cgp} S.Cecotti, L. Girardello and M.Porrati, Nucl.Phys B268 (1986) 295
\bibitem{b2} J.Bagger, Nucl. Phys Proc. Suppl. 52A (1997) 362, hep-th/9610022
\bibitem{fgp} S.Ferrara, L.Girardello and M.Porrati, Phys. Lett. B366 (1996) 155, hep-th/9510074;
Phys.Lett. B 376(1996) 275, hep-th/9512180
\bibitem{apt} I.Antoniadis, H.Partouche and J.R.Taylor, Phys.Lett. B372 (1996)83, hep-th/9512006
\bibitem{m} J.Michelson Nucl.Phys. B 495 (1996) 736, hep-th/9610151
\bibitem{tava}  T.R. Taylor and C. Vafa, Phys.Lett. B474 (2000) 130-137, hep-th/9912152
\bibitem{ma} P.Mayr, Nucl.Phys. B 59 (2001 99, hep-th/0003198
\bibitem{fks} S.Ferrara, R.Kallosh and A.Strominger, Phys. Rev. D52 (1995) 541, hep-th/9508072
\bibitem{ps} J.Polchinski and A.Strominger, Phys.lett. B 388 (1996) 736, hep-th/9510227
\bibitem{fk1} S.Ferrara and R. Kallosh, Phys.Rev. D 54 (1996) 1514, hep-th/9602136;
Phys.Rev. D 54 (1996) 1525, hep-th/9603090
\bibitem{fgk} S.Ferrara, G.Gibbons and R.Kallosh, Nucl. Phys. B500 (1997) 75, hep-th/9702103
\bibitem{kl} R.Kallosh and A.Linde, JHEP 0002 (2000) 005, hep-th/0001071
\bibitem{ckrrs} A.Chou, R.Kallosh, J.Rahmfeld, S.Rey, M.Shmakova
and W.K.Wong, Nucl.Phys. B508 (1997) 147, hep-th/9704142
\bibitem{gppz} L.Girardello, M.Petrini, M.Porrati and A.Zaffaroni,
JHEP 9812 (1998) 022, hep-th/9810126
\bibitem{pgpw} D.Z.Freedman, S.S.Gubser, K.Pilch and N.Warner, Adv.Theor.Math.Phys. 3 (1999) 363
hep-th/9904017
\bibitem{agmoo} O.Aharony,S.S.Gubser,J.Maldacena, H.Ooguri and
Y.Oz, Physics Reports 323 (2000) 183, hep-th/9905111
\bibitem{losw} A.Lukas, B.A.Ovrut, K.S.Stelle and D.Waldram,
Nucl.Phys. B552 (1999) 246, hep-th/9806051
\bibitem{bhlt}  K. Behrndt, C. Herrmann, J. Louis and S. Thomas, JHEP 0101 (2001) 011, hep-th/0008112
\bibitem{cd} A. Ceresole, G. Dall'Agata, Nucl.Phys. B585 (2000) 143-170, hep-th/0004111
\bibitem{gz} M.Gunaydin,and M.Zagermann, Phys.Rev. D63 (2001) 064023, hep-th/0004117
\bibitem{bc} K.Behrndt and M.Cvetic, hep-th/0101007
\bibitem{bgs} K.Behrndt, S.Gukov, M.Shmakova, hep-th/0101119
\bibitem{gs}  M.Gutperle, M. Spalinski, Nucl.Phys. B 528 (2001) 509, hep-th/0010192
\bibitem{bbs}  K.Becker, M.Becker and A.Strominger, Nucl.Phys.B456(1995) 130, hep-th/950715814
\bibitem{cfgv} E.Cremmer, S.Ferrara, L.Girardello and A.Van
Proeyen, Nucl. Phys. B212 (1983) 413
\bibitem{bw1} J. Bagger and E. Witten, Phys.Lett. B 115 (1982) 202
\bibitem{b1} J. Bagger, Nucl. Phys. B 211 (1983) 392
\bibitem{dlv} B.de Wit, P.G.Lauwers and A.Van Proeyen, Nucl.Phys.
B255 (1985) 569
\bibitem{a1} L. Andrianopoli, M. Bertolini, A. Ceresole, R. D'Auria, S. Ferrara, P. Fre'
 Nucl.Phys. B476 (1996) 397-417, hep-th/9603004 ;
  L. Andrianopoli, M. Bertolini, A. Ceresole, R. D'Auria, S. Ferrara, P. Fre' and T. Magri,
 J.Geom.Phys. 23 (1997) 111-189,  hep-th/9605032
 \bibitem{bw} J. Bagger and E. Witten, Nucl.Phys.B222(1983)1
 \bibitem{fs} S.Ferrara and S. Sabharwal, Nucl.Phys. B 232(1990) 317
 \bibitem{dff} R. D'Auria, S. Ferrara and P. Fr\'e, Nucl. Phys. B359 (1991) 705.
\bibitem{s} A.Strominger, Comm. Math. Phys. 133(1990) 163
\bibitem{cdf} L. Castellani, R. D'Auria and S. Ferrara,
Phys. Lett. 241B (1990) 57; Class. Quantum Grav. 7 (1990) 1767.
\bibitem{cdf1} L.Castellani, R.D'Auria and P.Fre`, "Supergravity and Superstrings: a Geometric Perspective",
Volume 2,World Scientific 1991
\bibitem{kls} R.Kallosh, A.Linde and M.Shmakova, JHEP 9911 (1999) 010, hep-th/9910021
\bibitem{al} D.V.Alekseevski, USSR Izvestija 9 (1975) 297
\bibitem{ga} K.Galiski, Comm.Math.Phys. 108 (1987) 117; Class. Quantum Grav. 9 (1992) 27
\bibitem{dv} B.de Wit and A.Van Proeyen, Int.Jou. of Mod.Phys.D3 (1994) 31,hep-th/9310067
\bibitem{dkv} B.de Wit, B.Klejn and S.Vandoren, Nucl.Phys. B 568 (2000) 475, hep-th/9909228
\bibitem{crtv} B.Craps, F.Roose, W.Troost and A. Van Proeyen, Nucl.phys. B 203 (1997) 565,hep-th/9703082
\bibitem{goi} A.Galperin, O.Ogievetsky and E.Ivanov, Ann. of Phys. 230 (1994) 201
\bibitem{cgf} S.Cecotti, S.Ferrara, and L.Girardello, Int.Jou.of
Mod.Phys. A4 (1989) 2475
\bibitem{bb} K.Becker and M. Becker, Nucl.Phys. B551 (1999) 102
\bibitem{cdfv} A. Ceresole, R. D'Auria, S. Ferrara, A. Van Proeyen, Nucl.Phys. B44(1995) 92, hep-th/9412200
\bibitem{cdfgkdv} E.Cremmer, C.Kounnas, S.Ferrara, J.P. Derendinger, B.de
Wit, L.Girardello and A.Van Proeyen, Nucl.Phys. B250 (1985) 385
\bibitem{adfft} L.Andrianopoli, R.D'Auria, S.Ferrara, P.Fre` and M.Trigiante, Nucl.Phys. B496 (1997) 617,
hep-th/9611114
\bibitem{cdkv} A.Ceresole, G.Dall'Agata, R.Kallosh and A.Van Proeyen, hep-th/0104056
\bibitem{Betal} M.Bianchi, O.DeWolfe, D.Z. Freedman and K.Pilch, hep-th/0009156
\end{thebibliography}
\end{document}